\documentclass[reprint,aps,showpacs,pra,nofootinbib,superscriptaddress]{revtex4-2}
\usepackage{color}
\usepackage[dvipsnames]{xcolor}
\usepackage{newlfont}
\usepackage{graphicx}
\usepackage{amssymb}
\usepackage{amsmath}
\usepackage[caption=false]{subfig}
\usepackage{bm}
\usepackage{verbatim}
\usepackage[latin1]{inputenc}
\usepackage{bbm}
\usepackage{multirow}
\usepackage[thinlines]{easytable}
\usepackage{braket}
\usepackage{amsthm}
\usepackage{bbold}
\usepackage{subfig}
\usepackage{mathrsfs}
\usepackage[varg]{txfonts} 
\usepackage{amsfonts}

\newcommand{\ba}{\begin{array}}
	\newcommand{\ea}{\end{array}}

\usepackage[pagebackref=false, colorlinks=true]{hyperref}
\definecolor{redish}{rgb}{0.7,0.2,0.0}  
\definecolor{bluish}{rgb}{0.2,0.5,0.8}
\hypersetup{linkcolor=red,          
	citecolor=teal,        
	filecolor=magenta,      
	urlcolor=blue}          

\begin{document}
\title{Generalised state space geometry in Hermitian and non-Hermitian quantum systems}
\author{Kunal Pal}
\email{kunal.pal@apctp.org} 
\affiliation{Asia Pacific Center for Theoretical Physics, Pohang 37673, Republic of Korea}

\begin{abstract}
One of the key features of information geometry in the classical setting is the existence of a metric structure and a family of connections on the space of probability distributions. The uniqueness of the Fisher--Rao metric and the duality of these connections is at the heart of classical information geometry. However, these features do not carry over straightforwardly to quantum systems, where a Hermitian inner product structure on the Hilbert space induces a metric on the complex projective space of pure states---the Fubini-Study tensor, which is preserved under the unitary evolution. In this work, we explore how modifying the Hermitian tensor structure on the projective space may affect the geometry of pure quantum states, and whether such generalisations can be used to define dual connections with a direct correspondence to classical probability distribution functions, modified by the presence of a non-trivial phase. We show that it is indeed possible to construct a family of connections that are dual to each other in a generalised sense with respect to the real-valued sector of the Fubini--Study tensor.
Using this biorthogonal formalism, we systematically classify the four types of tensors that can arise when the dynamics of a quantum system are governed by a non-Hermitian Hamiltonian, identifying both the complex-valued metric and the Berry curvature. Finally, we elucidate the role of the metric in a quantum natural gradient descent optimisation problem, generalised to the non-Hermitian case for a suitable choice of cost function.

\end{abstract}
\maketitle


\section{Introduction}
In classical information geometry, it is possible to ascribe a unique differential geometric structure on the space of probability distributions by means of the Fisher-Rao metric and the family of $\alpha$-connections, which provide a notion of distance and parallel transport on these generally curved spaces \cite{Amari2000methods}. A collection of probability distributions \( P(x; \theta) \) for a random variable \( x \),  parametrised by a set of continuous parameters \( \theta\), can be most conveniently thought of as a differential manifold, which is equipped with the classical information metric that quantifies the notion of distance in terms of statistical distinguishability: two distributions are considered distant if they can be reliably differentiated with only a small number of observations of \( x \). The classical Fisher-Rao metric can then be written for a well-defined and normalised probability distribution as 
\begin{equation}
g^{FR}_{ij}=\mathcal{E}_{p}\Big[\partial_{i} \ln  P(x;\theta)\partial_{j}\ln  P(x;\theta)\Big]~.
\label{Fishermetric}
\end{equation}

Here and in subsequent discussions, we use $\mathcal{E}_{p}[]$ to represent the statistical average of a quantity with respect to the probability distribution function under consideration, and the partial derivatives are with respect to the parameters $\theta_{i}$.
The classical information metric tensor and the $\alpha$-family of connections can be obtained from a consistent expansion of the one-parameter family of the so-called divergence functionals. Importantly, defining the metric and the family of connections gives rise to a duality between the $\alpha$ and $-\alpha$ connections with respect to the metric \cite{Chentsov2000statistical, Amari2012differential}. 

The classical formulation of information geometry has been widely applied to various fields, particularly in classical statistical systems, to understand phenomena ranging from phase transitions to the emergence of chaotic properties. The Fisher information metric defines a Riemannian structure on parameter spaces of probability distributions, enabling the study of phase transitions through curvature singularities \cite{Ruppeinerrev}. In equilibrium thermodynamics, the information geometric formulation and the related Ruppeiner geometrical picture have been used to investigate the critical behaviour and thermodynamic stability of fluids, black holes, and spin systems \cite{Janyszek, Johnston, Dolan, Brody_2009, Crooks, TS1, TS2, Cafaro2007, Aaman, TS3, Ruppeiner2012, paltorsion}. The scalar curvature derived from the Fisher metric often encodes information about the interaction strength and the correlation length in many-body systems. This geometric perspective complements traditional approaches by linking thermodynamic fluctuations to an underlying statistical manifold structure \cite{RuppeinerAJP}.  \footnote{The references cited above represent only a very selective section of works that have appeared over the years; for a complete history and references, we refer the reader to the excellent reviews \cite{Ruppeinerrev, Brody_2009}.}

For quantum systems, on the other hand, parametrised by a set of parameters, the notion of distance between quantum states can be defined in the space of quantum states or density matrices \cite{Provost1980riemannian}. On the complex projective space of the pure quantum states,  a Hermitian tensor structure can be written down, in accordance with the inner product on the Hilbert space. This is the natural Fubini-Study (FS) tensor structure for a quantum state $\Psi$, which we have assumed to be normalised to unity and is by construction invariant under $U(1)$ transformations \cite{Brodygeometric}.
Then the pull-back of the FS tensor of the complex projective space to the parameter manifold is essentially what is known as the quantum geometric tensor (QGT), which can be written in the real coordinates, parameterising the pure quantum state $\{\theta_{i}\}$ as \cite{Ashtekar, Kibble1979, Braunstein94, Field1999geometry, Anandan1991} as 
\begin{equation} 
FS_{ij}=\braket{\partial_{i}{\Psi(\theta)}|\partial_{j}\Psi(\theta)}-\braket{\partial_{i}\Psi(\theta)|\Psi(\theta)}\braket{\Psi(\theta)|\partial_{j}\Psi(\theta)}~.
\label{tensorstructure}
\end{equation}
 
Starting from this tensor  on the complex projective space of pure states, it was shown in \cite{Facchi2010classical} that it is possible to write the real and symmetric part of the FS tensor in explicit coordinate notation as 
\begin{widetext}
\begin{equation}
\begin{split}
    g^{FS}_{ij}=\frac{1}{4}\mathcal{E}_{p}\Big[\partial_{i}\ln  {P(x;\theta)}\partial_{j}\ln  {P(x;\theta)}\Big]+ \mathcal{E}_{p}\Big[ \partial_{i}\phi(x;\theta)\partial_{j}\phi(x;\theta)\Big]-\mathcal{E}_{p}\Big[\partial_{i}\phi(x;\theta)\Big]\mathcal{E}_{p}\Big[\partial_{j}\phi(x;\theta)\Big]~,
\label{FSmetric}
\end{split}
\end{equation}
\end{widetext}
known as the quantum metric tensor (QMT).
On the other hand, the closed $2$-form in the projective space, given by the imaginary antisymmetric tensor
\begin{equation}
\omega_{ij}=\frac{i}{2}\mathcal{E}_{p}\Big[\partial_{i}\ln{P(x;\theta)} \partial_{j}\phi(x;\theta)-\partial_{j}\ln  P(x;\theta)\partial_{i}\phi(x; \theta)\Big]~,
\label{berrycuravture}
\end{equation}
and known as the Berry curvature in the Physics literature, defines a symplectic structure on the manifold. 
Here, we have used the polar decomposition of the position-space wave function in terms of the two real functions as;
\begin{equation}
\Psi(x;\theta)=\sqrt{P(x;\theta)}e^{i\phi(x;\theta)},
\end{equation}
 parametrised by the set of $n$ parameters $\{\theta_{i}\}={\theta_{1}, \theta_{2},\dots\theta_{n}}$. The wave function and consequently the two functions $P(x;\theta)$, and $\phi(x;\theta)$ are assumed to be smooth and differentiable functions everywhere on the parameter manifold. It should be noted that the form of a metric tensor \eqref{FSmetric} introduced in \cite{Facchi2010classical}, even for a classical PDF and a quantum state in position space representation with the same probability amplitude $P(x;\theta)$ will have a different metric structure on the parameter manifold due to the presence of a non-trivial phase of the wavefunction, which is the explicit manifestation of the quantum nature of the system. The Berry curvature can also be thought of as the field strength tensor associated with the Berry connections $A_{i}=i\braket{\Psi(\theta)|\partial_{i}\Psi(\theta)}$, as $F_{ij}=\partial_{i}A_{j}-\partial_{j}A_{i}$, which is manifestly antisymmetric in the two indices \cite{Berry1984}.

The fact that the natural, flat Hermitian inner product on the Hilbert space of states induces a convenient notion of `distance' between two nearby states on the Projective Hilbert space can be perhaps best seen by considering the overlap of two states separated by a small parametric distance of the form $\braket{\Psi(\theta+\delta\theta)-\Psi(\theta)|\Psi(\theta+\delta\theta)-\Psi(\theta)}$, as was done by Provost and Vallee in \cite{Provost1980riemannian}. Expanding this overlap of states and demanding the invariance of the metric under a global phase transformation, it is possible to obtain what is now known as the Provost-Vallee metric (PV), which is essentially identical in form with \eqref{tensorstructure}. 

The QGT plays a central role in understanding the structure and dynamics of quantum many-body systems. It encodes both the quantum metric, which measures the infinitesimal distance between nearby ground states, and the Berry curvature, which captures geometric and topological properties. In many-body systems, the QGT has been used to analyse the response of ground states to changes in external parameters, enabling the study of phase transitions from a geometric viewpoint \cite{Zanardi2006, Zanardi2007, Dey, Streleck, Kolodrubetz}. The quantum metric component directly relates to fidelity susceptibility, which shows critical scaling behaviour near quantum phase transitions \cite{venuti2007quantum}. This allows for the detection of phase transitions without invoking symmetry-breaking or local order parameters. The Berry curvature component, meanwhile, plays a crucial role in characterising topological phases and computing topological invariants such as the Chern number \cite{resta2011insulating, Ozawa, Mera21}. In the context of band theory, the Berry curvature acts as a local `effective magnetic field' in momentum space and is central to understanding topological properties of Bloch bands \cite{Xiaorev}, where it underlies quantised responses such as the anomalous Hall effect and the Chern number. Over the last decade or so, this differential geometric formulation has been used widely in related areas for more detailed discussions and complete references, we will point to the review articles \cite{gu2010fidelity, Lambert}.


In this context, it is a natural question to ask how far the standard picture of classical $\alpha$-connection of information geometry can be extended for the quantum mechanical case, and what physical insights can be obtained from such a construction, in particular for quantum many-body systems. One of the key results to this end was proved in \cite{Morozova1991markov}, where it was shown that the uniqueness  of the Fisher-Rao metric is no longer valid for a generic quantum system \cite{Petz1996}. This discovery spurred a surge of research activities in the quantum information geometry community about the structure of quantum $\alpha$-connection and various formulations of divergence functional \cite{Hasegawa1993alpha, Hasegawa1995, Hasegawa1997, Jenvcova2001geometry, Petz:1999, Naudts18, Naudts24, Molitor}. However, a proper review of the literature, justifying the richness of the subject, is beyond the scope of the present paper, and we will point to the recent rigorous review of various aspects of quantum information geometry \cite{Ciaglia2023}. 

Our primary aim in this work is to explore how the structure of the inner product on the Hilbert space changes the QGT induced on the parameter submanifold. Instead of directly working with the pure quantum state $\Psi(x; \theta)$ we will, in most cases use it, as written in terms of two real functions $P(x;\theta)$ and $\phi(x;\theta)$, in the position space representation, which of course encodes the same information as that of the PV metric, but in our opinion has more transparency, in particular, when exploring the different connections on the manifold of pure states from a quantum many-body theory perspective, as we will argue in subsequent sections. We will explore how far the standard metric (eq. \eqref{FSmetric}) and the Berry curvature (eq. \eqref{berrycuravture}) constructions can be extended to `mimic' the various features of classical information geometry, from more of a `phenomenological' point of view rather than starting from a first-principle extension of information geometry for quantum systems. 

To this end, first, we will discuss some basic features associated with the metric connection of the QMT and what kind of structures we should expect when trying to generalise this for a parametric distribution.
Then we will show why it is not possible to generalise the overlap integral for a quantum state using the standard Hermitian inner product except in the special case $\alpha=0$, while satisfying the normalisation condition. Motivated by this observation, we will next introduce a class of tensor structures on the space of wave functions that are not \textit{manifestly} Hermitian and require the $\alpha$-representation in terms of two different functions, which can be obtained from a consistent expansion of the biorthogonal overlap integral. Then we will elucidate the role of the QMT, Berry curvature and the role of two conjugate connections, the last two can be shown to satisfy an extended duality in this case.  In particular, we will discuss why the real parts of the two gauge-invariant connections for these two functions satisfy a $\pm\alpha$ connection duality, very similar to the classical case. Finally, we will show how the machinery developed in this case can be directly carried over to quantum systems, where the dynamics are governed by a non-Hermitian Hamiltonian, giving rise to a consistent classification of four types of tensor structures that can appear. The analogue of the Berry curvature can then be obtained from the complex-valued Berry connection, and the role of the corresponding QMT is elucidated by considering a quantum natural gradient descent optimisation problem for the non-Hermitian systems.


\section{Connections on the parameter manifold}
\label{symmetricompatibleconnection}
In this section, we will explore the QMT and various geometrical quantities associated with the metric, and also those that are independent of the metric on the parameter manifold. We will start with the metric connection derived from the FS metric \eqref{FSmetric}, and then we will provide a possible phenomenological extension to include the non-metric connection more common in classical information geometry \cite{Amari2000methods}. This section will be used to discuss some simple constructions for pure-state-based analogues of classical information geometry, and in doing so, we will establish the notation used throughout the rest of the paper.

\subsection{Symmetric and metric-compatible connection}
\label{metricocnnection}
Let us first consider the 
natural connection on the parameter manifold that preserves both the metric and the symplectic structure, which is the standard metric connection, constructed from the metric tensor \eqref{FSmetric}. This does not transform like a tensor and is of the following form (in the real coordinates $\theta_{i}$'s)
\begin{widetext}
\begin{equation}
\begin{split}
\Gamma^{(c)}_{ij,k}=\text{Re}\Bigg[\braket{\partial_{i}\partial_{j}\Psi|\partial_{k}\Psi}-\braket{\partial_{i}\partial_{j}\Psi|\Psi}\braket{\Psi|\partial_{k}\Psi}
-\Bigg(\braket{\partial_{i}\Psi|\Psi}\braket{\partial_{j}\Psi|\partial_{k}\Psi}+\braket{\partial_{j}\Psi|\Psi}\braket{\partial_{i}\Psi|\partial_{k}\Psi}\Bigg)
\Bigg]~,
\label{GammacfromQFI}
\end{split}
\end{equation}
\end{widetext}
where we have indicated $\frac{\partial^2 f}{\partial\theta^{i}\partial\theta^{j}}$ as $\partial_{i}\partial_{j}f$.
The symmetry in the first two indices is manifest in $\Gamma^{(c)}_{ij,k}$, and the Hermitian nature of the FS tensor implies the metric-connection is real. Note that this is a symmetric (in the first two indices) rank-3 object built directly from the QMT, which corresponds to the uncontracted form of the Levi-Civita connection components. When needed, the standard Christoffel symbols can be recovered by contracting with the inverse metric, provided that the inverse exists.

Then, using the polar decomposition of the wavefunction in the position space, we can obtain the following form,
\begin{widetext}
\begin{equation}
\begin{split}
\Gamma^{(c)}_{ij,k}=\frac{1}{2}\Bigg(\frac{1}{4}\mathcal{E}_{p}\Big[\underbrace{2\partial_{i}\partial_{j}l_{\theta} \partial_{k}l_{\theta}+\partial_{i}l_{\theta}\partial_{j}l_{\theta}\partial_{k}l_{\theta}}_{\text{Classical metric connection}}\Big]+
\mathcal{E}_{p}\Big[2\partial_{i}\partial_{j}\phi\partial_{k} \phi+2\partial_{(i}l_{\theta}\partial_{j)}\phi\partial_{k} \phi
-\partial_{i}\phi\partial_{j}\phi\partial_{k}l_{\theta}\Big]-\\
\Bigg\{2\mathcal{E}_{p}\Big[\partial_{i} \partial_{j} \phi\Big]\mathcal{E}_{p}\Big[\partial_{k} \phi\Big]+2
\mathcal{E}_{p}\Big[ \partial_{(i} \phi\partial_{j)} l_{\theta} \Big]\mathcal{E}_{p}\Big[\partial_{k} \phi\Big]+2
\mathcal{E}_{p}\Big[\partial_{i}\phi\Big]\mathcal{E}_{p}\Big[ \partial_{[k} \phi\partial_{j]} l_{\theta}\Big]+2
\mathcal{E}_{p}\Big[ \partial_{j} \phi\Big]\mathcal{E}_{p}\Big[ \partial_{[k} \phi\partial_{i]} l_{\theta}\Big]
\Bigg\}\Bigg)~.
\label{metricicnnevtionPphi}
\end{split}
\end{equation}
\end{widetext}
From now on, we will use the notation $l_{\theta}=\ln{P}$, and the dependence on $x,\theta$ will be suppressed. Also, we have denoted the symmetrisation over two indices as $(ij)$ and antisymmetrisation as $[ij]$. As can be seen, the connection is symmetric in the first two indices $(ij)$, which confirms the torsion tensor for this connection identically vanishes everywhere on the manifold and the covariant derivative of the FS metric with respect to this connection is $\nabla_{k}g^{FS}_{ij}=0$, as is expected. Expressing the connection in terms of the polar decomposition \((P, \phi)\) of the wavefunction allows a transparent separation between classical (amplitude-based) and genuinely quantum (phase-based) contributions to geometry. In particular, the terms involving \(l_{\theta} = \ln P\) alone encode the classical Fisher information, while those involving \(\phi\) capture purely quantum features such as the Berry curvature. This formulation not only clarifies the relation to classical information geometry but also provides a natural setting for phenomenological generalisations, especially in systems where position-space representations are more tractable.
 
\subsection{Symmetric and non-metric connections on the manifold}
The connection we obtained in the last section, though natural on the state manifold, is not very useful for constructing an affine-coordinate system for a given non-trivial metric. This motivates us to construct a one-parameter family of connections, which are generally not metric connections, so they have a `non-metric' contribution.  To this end, we consider the following map $\Gamma^{(\alpha)}_{ij,k}$, which maps each point of $P(x;\theta)$ and $\phi(x;\theta)$ to the corresponding value, for any real number $\alpha$ \cite{Amari2000methods}:
\begin{equation}
\Gamma^{(\alpha)}_{ij,k}=\Gamma^{(c)}_{ij,k}+N_{ij,k}~. 
\label{Alphasconnection}
\end{equation}
Here, the contribution from the non-metric part of the connection is given by
\begin{widetext}
\begin{equation}
\begin{split}
    N_{ij,k}=-\frac{\alpha}{2}\cdot\frac{1}{4}\mathcal{E}_{p}\Big[\underbrace{\partial_{i}l_{\theta}\partial_{j}l_{\theta}\partial_{k}l_{\theta}}_{\text{Classical Non-metricity}}\Big]-\frac{\alpha}{2}\mathcal{E}_{p}\Big[2\partial_{i}\partial_{j}\phi\partial_{k} \phi+2\partial_{(i}\phi\partial_{j)}l_{\theta}\partial_{k}\phi-\partial_{i}\phi\partial_{j}\phi\partial_{k}l_{\theta}\Big]+\\
\frac{\alpha}{2}\Bigg(2\mathcal{E}_{p}\Big[\partial_{i} \partial_{j} \phi\Big]\mathcal{E}_{p}\Big[\partial_{k} \phi\Big]+2\mathcal{E}_{p}\Big[\partial_{(i}\phi\partial_{j)}l_{\theta}\Big]\mathcal{E}_{p}\Big[\partial_{k}\phi\Big]+2
\mathcal{E}_{p}\Big[\partial_{i}\phi\Big]\mathcal{E}_{p}\Big[\partial_{[j}l_{\theta}\partial_{k]}\phi\Big]+2
\mathcal{E}_{p}\Big[\partial_{j}\phi\Big]\mathcal{E}_{p}\Big[\partial_{[i}l_{\theta}\partial_{k]}\phi\Big]\Bigg)~.
\label{Nonmetricitty}
\end{split}
\end{equation}
\end{widetext}
As can be clearly seen, the tensor $N_{ij,k}$ is symmetric in the first two indices $(i,j)$ and consequently so is the full connection $\Gamma^{(\alpha)}_{ij,k}$, which we will call the $\alpha$-connection, since in this context it looks like a modified version of the classical $\alpha$-connection, following the standard terminology of \cite{Amari2000methods}. Also, it should be noted that the full connection is not a metric connection for a general $\alpha\neq 0$; on the other hand, for $\alpha=0$, this reduces to the standard metric connection on the manifold with respect to the quantum metric tensor. For states with a trivial phase factor, on the other hand, the non-metricity tensor (as well as the total connection ) reduces to the  $\alpha$-connections of classical setting, with a factor of $1/4$ \cite{Amari2000methods}. It can also be checked that the non-metricity part of the tensor $N_{ij,k}$ is invariant under a phase redefinition, and therefore is a physically valid contribution to the geometry.
At this point, we stress the form of the nonmetric connection that we obtain in eq. \eqref{Nonmetricitty} is not unique, as the conditions we have imposed to obtain this are not restrictive, and hence it is possible to have other forms of nonmetricity tensor consistent with the requirement. A proper form of connection is only possible to obtain from a consistent expansion of the divergence functional. However, it is known that the metric and connections in the quantum case are not unique, unlike in the classical case \cite{Chentsov2000statistical, Jenvcova2001geometry, Jencova2003}, and the precise relationship between such a connection and that obtained here remains to be explored. Our point in doing this exercise is to construct an affine coordinate system for a class of PDFs, which has the potential to simplify many of the subsequent analyses, and how that can be explained from a geometric point of view remains to be seen.
 
\subsubsection{Example: Probability distribution function belonging to the classical exponential family}
Let us assume that the form of the position space pure state wavefunction is given by,
\begin{equation}
\Psi(x;\theta)=\exp\Bigg[\frac{1}{2}\Bigg({C(x)+\sum^{n}_{j=1}\theta^{j}\Big(F_{j}(x)+i G_{j}(x)\Big)-\psi(\theta)}\Bigg)\Bigg].
\label{Exponentialwavefunction}
\end{equation}
Here, $C(x), F_{j}(x), G_{j}(x) $ are real-valued functions solely of $x$ and $\psi(\theta)$ is only a function of $\theta$ and we have ignored a global phase factor, which will not contribute anything physically meaningful. Note that the set of functions $\{F_{1},..F_{n},1\}$ has to be linearly independent for the map $\theta \rightarrow P(x;\theta)$ to be one-to-one \cite{Amari2000methods}.
The corresponding form of the probability distribution function $P(x;\theta)$ belongs to the so-called `exponential family' of the PDF \footnote{In our current context, this is a formal expression of the position space wave function, in contrast, the standard quantum exponential family is represented in the space of density matrices as $\rho(\theta)=\exp({\theta^{i}\mathcal{O}}_{i}-\psi(\theta))$, for a set of Hermitian operator $\mathcal{O}$, representing the observables \cite{Hasegawa1993alpha, Nakamura2022scalar, Hasegawa1995}.}. Then from the normalisation of the wavefunction, we can get the expression of the function $ \psi(\theta)$, 
\begin{equation}
    \psi(\theta)=\ln\Bigg(\int \exp{\Big[C(x)+\sum^{n}_{j=1}\theta^{j}F_{j}(x)\Big]} dx\Bigg)~.
\end{equation}
Then the FS metric for this wavefunction can be written in terms of the functions $\psi(\theta)$ and $G_{i}$'s as  
\begin{equation}
    g^{FS}_{ij}=\frac{1}{4}\partial_{i}\partial_{j}\psi(\theta)+\mathcal{E}_{p}\Big[G_{i}(x)G_{j}(x)\Big]-\mathcal{E}_{p}\Big[G_{i}(x)\Big]\mathcal{E}_{p}\Big[G_{j}(x)\Big]~, 
\end{equation}
which, in general, is different from the simple classical Fisher information metric having the same exponential class of PDF. The second term in the Fubini-Study metric subtracts the covariance of the quantum phase, isolating only the physically relevant, gauge-invariant contributions: 
\begin{equation}
   g^{FS}_{ij} = \frac{1}{4} \partial_i \partial_j \psi(\theta) + \mathrm{Cov}_P[G_i(x), G_j(x)]~. 
\end{equation}

On the other hand, the closed $2$-form on the complex projective space is
\begin{widetext}
  \begin{equation}
\omega_{ij}=\mathcal{E}_{p}\Big[F_{i}(x)G_{j}(x)-F_{j}(x)G_{i}(x)\Big]-\partial_{i}\psi(\theta)\mathcal{E}_{p}\Big[G_{j}(x)\Big]-\partial_{j}\psi(\theta)\mathcal{E}_{p}\Big[G_{i}(x)\Big]~.
\end{equation}  
\end{widetext}

\subsubsection{Metric connection: Exponential family}
For the wavefunction which in the position space representation $\Psi(x;\theta)$ has the form of \eqref{Exponentialwavefunction}, we can obtain the form of the metric compatible connection, which in this case has the form 
\begin{widetext}
\begin{equation}
    \begin{split}
    \Gamma^{(c) exp}_{ij,k}=\frac{1}{2}\Bigg(-\frac{1}{4}\mathcal{E}_{p}\Big[2\partial_{i}\partial_{j}\psi(\theta)A_{k}\Big]+\mathcal{E}_{p}\Big[A_{i}A_{j}A_{k}\Big]+\mathcal{E}_{p}\Bigg[\frac{G_{k}}{2}G_{(i}A_{j)}-\frac{1}{4}G_{i}G_{j}A_{k}\Bigg]
    -\Big(\mathcal{E}_{p}\Big[\frac{G_{k}}{2}\Big]\mathcal{E}_{p}\Big[G_{(i}A_{j)}\Big]+\\
    \mathcal{E}_{p}\Big[\frac{G_{i}}{2}\Big]\mathcal{E}_{p}\Big[A_{[j}G_{k]}\Big]+\mathcal{E}_{p}\Big[\frac{G_{j}}{2}\Big]\mathcal{E}_{p}\Big[A_{[i}G_{k]}\Big]\Big)
    \Bigg)~,
\end{split}
\label{Exponentialmetricconnection}
\end{equation}
\end{widetext}
which, as expected, does not vanish identically for the exponential family. Here we have defined the $A_{i}=F_{i}(x)-\partial_{i}\psi(\theta)$ for clarity. The metric-compatible connection includes additional terms beyond the classical exponential connection, arising from non-trivial covariances between amplitude and phase, and ensures compatibility with the full Fubini-Study structure.

\subsubsection{Non-metric connection: Exponential family}
In this section, we will introduce a generalised version of the $\alpha$-connection that is not compatible with the FS metric for the exponential family of PDFs. To this end, we introduce the one-parameter dependent non-metricity tensor of the following form for the exponential family 
\begin{widetext}
\begin{equation}
\begin{split}
    N^{\text{exp}}_{ij,k}=-\frac{\alpha}{2}\frac{1}{4}\mathcal{E}_{p}\Big[A_{i}A_{j}A_{k}\Big]-\frac{\alpha}{2}\mathcal{E}_{p}\Big[\frac{G_{k}}{2}G_{(i}A_{j)}-\frac{1}{4}G_{i}G_{j}A_{k}\Big]
    +\frac{\alpha}{2} \Bigg(\mathcal{E}_{p}\Big[\frac{G_{k}}{2}\Big]\mathcal{E}_{p}\Big[G_{(i}A_{j)}\Big]+
    \mathcal{E}_{p}\Big[\frac{G_{i}}{2}\Big]\mathcal{E}_{p}\Big[A_{[j}G_{k]}\Big]+\mathcal{E}_{p}\Big[\frac{G_{j}}{2}\Big]\mathcal{E}_{p}\Big[A_{[i}G_{k]}\Big]\Bigg)~.
\end{split}
\end{equation}
\end{widetext}
As can be seen from the form of the connection coefficients, for the wavefunction of the form \eqref{Exponentialwavefunction}, all the components of $\Gamma^{(1)}_{ij,k}$ vanish, which is reminiscent of the corresponding case of $1$-connection for the classical exponential family of PDFs \cite{Amari2000methods}. Thus the coordinate system $\{\theta_{i}\}$ is an $1$-affine coordinate system, and the parameter manifold $S$, is $1$-flat. Even though this looks promising, a proper form of non-metric connection will be derived later from a consistent expansion of the overlap integral.

\subsection{Connections from the Fisher information}
\label{connectionfromQFI}
The quantum counterpart of the classical Fisher information, known as the quantum Fisher information (QFI), introduced by Helstrom \cite{Helstrom1967minimum}, provides an upper bound on the classical Fisher information. For the pure state manifold, it was shown in \cite{Facchi2010classical}, by expanding the QFI, written in terms of the symmetric logarithmic derivative, how to obtain the tensor field on the complex projective space that incorporates the Fubini-Study metric and the antisymmetric $g_{ij}$, closed 2-form $\omega_{ij}$. In this section, our aim is to show how we can obtain the form of both the metric-compatible connection by expanding the suitably generalised QFI in a similar spirit to \cite{Facchi2010classical}. 
\subsubsection{Metric compatible connection}
We start by writing the Hermitian Fubini-Study tensor in an explicit coordinate-dependent form with respect to the real coordinates; we have 
\begin{equation}
    \mathcal{F}_{g}:=\frac{1}{4}\text{Tr}\Big[\rho\partial_{i}\rho\partial_{j}\rho+\partial_{j}\rho\partial_{i}\rho\rho+\rho\partial_{j}\rho\partial_{i}\rho+\partial_{i}\rho\partial_{j}\rho\rho\Big]
\label{FSmetricfromQFI}
\end{equation}
for a pure state density matrix of the form $\rho=\ket{\Psi}\bra{\Psi}$. Here, the symbol $:=$ indicates that we have to substitute the form of the density matrix and have to take the trace to arrive at the form of the metric, which is explicitly real and symmetric. As was shown in \cite{Facchi2010classical} for the FS metric, in a straightforward way we can similarly obtain the following form of the metric-compatible form of connection coefficients for the pure state density matrix from a generalised QFI-like quantity of the form 
\begin{widetext}
\begin{equation}
\begin{split}
\mathcal{F}_{\Gamma^{(c)}}:=\frac{1}{4} \text{Tr}\Big[\rho\partial_{i}\partial_{j}\rho\partial_{k}\rho+\rho\partial_{k}\rho\partial_{i}\partial_{j}\rho+\partial_{i}\partial_{j}\rho\partial_{k}\rho\rho+\partial_{k}\rho\partial_{i}\partial_{j}\rho\rho+
\partial_{i}\rho\partial_{k}\rho\partial_{j}\rho+\partial_{j}\rho\partial_{k}\rho\partial_{i}\rho\Big]~.
\label{QFIgammac}
\end{split}
\end{equation}
\end{widetext}
Here, the trace operation has to be taken with respect to a fixed basis on the Hilbert space, and we have written each term in a way such that the real and symmetric (in indices $i,j$) nature of this quantity is evident. In obtaining \eqref{GammacfromQFI} from \eqref{QFIgammac} we have used $\text{Tr}[\rho]=1$, the fact that the state $\ket{\Psi}$ is properly normalised. However, from the symmetries of the non-metricity part of the connection, it is clear that it cannot be directly obtained from the symmetric QFI. Hence, they encode non-QFI contributions of the quantum states to the geometry, which is not represented by the FS metric.


\section{$\alpha$-connections for the quantum parameter manifold}
\label{alphaconnectionsection}
In this section, we will briefly review some basic aspects of the $\alpha$-representations of the tangent space of the space spanned by functions and the corresponding metric and the $\alpha$-connection structure (see \cite{Amari2000methods} for more details). To this end, it is customary to define a one-parameter family of functions of the form 
\begin{equation}
\begin{split}
    F_{\alpha}(x)=\frac{2}{1-\alpha}x^{\frac{1-\alpha}{2}}, \hspace {3mm}\text{For} \hspace {1mm} \alpha \neq 1\\
    =\log x \hspace {3mm}\text{For}  \hspace {1mm} \alpha= 1.~~~~~~~~~~~~
\end{split}
\end{equation}
Then for the space of normalised PDF $P(x,\theta)$, the metric tensor is defined as \cite{Amari2000methods}
\begin{equation}
    g^{(\alpha)}_{ij}(\theta)= \int \partial_{i}l_{\alpha}(x;\theta)\partial_{j}l_{-\alpha}(x;\theta) ~dx,
\label{alphametric}
\end{equation}
Where we have defined the functional $l_{\alpha}(x;\theta)=F_{\alpha}(P(x;\theta))$. This definition reduces to the well-known version of the Fisher-Rao information metric for $\alpha=0$. Then the corresponding set of $\alpha$-connections are defined as follows:
\begin{equation}
    \Gamma^{(\alpha)}_{ij,k}(\theta) =\int \partial_{i}\partial_{j}l_{\alpha}(x;\theta)\partial_{k}l_{-\alpha}(x;\theta) dx~.
\end{equation}
Using the form of the derivatives explicitly, we obtain
\begin{equation}
    \Gamma^{(\alpha)}_{ij,k}(\theta) =\int \Big(\partial_{i}\partial_{j}l_{\theta}\partial_{k}l_{\theta}+\frac{1-\alpha}{2} \partial_{i}l_{\theta}\partial_{j}l_{\theta}\partial_{k}l_{\theta}\Big) P(x;\theta)dx~.
\end{equation}
From this expression, it is easy to check that for the exponential family defined above, the $1$-connection  identically vanishes. Thus, the set of parameter space $\{\theta^{i}\}$ is called an affine coordinate system in this case.
This class of symmetric connections satisfies the well-known duality for the $\alpha$ and the $-\alpha$ family, namely,
\begin{equation}
\partial_{k}g^{\alpha}_{ij}=\Gamma^{(\alpha)}_{ik,j}+\Gamma^{(-\alpha)}_{jk,i},
\label{classicalIGduality}
\end{equation}
with 
\begin{equation}
    (\Gamma^{(-\alpha)}_{ij,k})^{*}=\Gamma^{(\alpha)}_{ij,k}~,
\end{equation}
where $*$ represents the duality operation.
Both the metric and the $\alpha$-connection can be systematically obtained by expanding the so-called $\alpha$-Divergence $D^{(\alpha)}\Big(P(x;\theta)||Q(x;\theta^{\prime})\Big)$, which measures the distinguishability of two PDFs $P(x;\theta)$ and $Q(x;\theta^{\prime})$ with slightly different parameter values. 

Our aim here primarily is to obtain a generalisation of the Fubini-Study metric and the corresponding connection to the case of the general information metric on the quantum parameter manifold for pure states described by a wavefunction $\Psi(x;\theta)$, which smoothly depends on a set of parameters $\{\theta^{i}\}$. We construct a one-parameter generalisation of the FS metric that we continue to call the $\alpha$-FS metric and the corresponding connection as the $\alpha$-FS connection, even though the connection will not conserve the $\alpha$-FS metric in general. We will build this structure based on the following checks listed as follows: 
\begin{itemize}
    \item (Check-1) For $\alpha=0$, the constructed metric and the $\alpha$ connection should be reduced to the corresponding FS metric and the FS connection, which is also a metric connection in this case.
    \item (Check-2) If the phase $\phi(x;\theta)$ of the pure state is trivial, then the $\alpha$-FS metric and the $\alpha$-FS connection should reduce to the $\alpha$-metric and the $\alpha$-connection of the classical information geometry.
\end{itemize}
We will also check if the following structures of classical information geometry are preserved or not and how far we can go without violating one or more of the following conditions;
\begin{itemize}
    \item (Condition-1) If these sets can be obtained from a generalised form of QFI similar to that introduced in section \ref{connectionfromQFI}. 
   
    \item (Condition-2) Are the $\alpha$ -FS connections obtained satisfying the duality structure of the classical IG, namely, eq. \eqref{classicalIGduality}?
\end{itemize}
Our approach is primarily based on generalising the structures of the FS metric for general one-parameter cases, which, of course, leaves room for huge degeneracies, as it is possible to find indefinitely many representations that reduce to the FS metric for the case $\alpha=0$. Nevertheless, we will introduce as minimal deformations as possible to obtain these structures on the quantum parameter manifold and check if those conditions listed above are satisfied or not. A more systematic approach should be based on using the generalised $\alpha$-divergence for the quantum case as a starting point. Such constructions have a long history and were pursued, for example, in \cite{Hasegawa1993alpha, Jenvcova2001geometry}; however, we will not follow such approaches in the present work.

\subsection{Case-1: Symmetric generalisation of the Fubini-Study metric}
Let us start with the simplest possible generalisation of the Fubini-Study tensor structure on the complex projective space of the pure quantum state to the one-parameter family, parametrised by $\alpha$. We propose to study the following tensor structure on the manifold  
\begin{equation}
FS^{(\alpha)}_{ij}=\braket{\partial_{i}{l_{(\alpha)}}|\partial_{j}l_{(-\alpha)}}-\braket{\partial_{i}l_{(\alpha)}|l_{(-\alpha)}}\braket{l_{(\alpha)}|\partial_{j}l_{(-\alpha)}}~.
\label{generaltensorstructure1}
\end{equation}
Here we have used the Dirac notation for $l_{\alpha}$, which, when written in the position representations (collectively denoted as $x$) are assumed to be of the form
\begin{equation}
    l_{\alpha}(x;\theta)=\frac{1}{1-\alpha}(\Psi(x;\theta))^{1-\alpha}=\frac{P(x;\theta)^{\frac{1-\alpha}{2}}}{1-\alpha}e^{i(1-\alpha)\phi(x;\theta)}~,
\end{equation}
where we have used the polar representation of the wave function in the position space. From now on, we will consider particularly the cases where $\alpha\neq1$.  As we will see in the sequel, this naive generalisation, directly motivated by the $\alpha$-representation  of the classical case, will not provide a consistent description of the geometry, as in the presence of non-trivial phase, they are not normalised with respect to each other. Nevertheless, we have chosen to perform the exercise in the rest of the section to stress why for quantum systems, it is natural to consider non-Hermitian structures for $\alpha$-representations, how this will enrich the metric structure on the manifold, and to establish some notations along the way.

First, note that the tensor structure assumed above is not manifestly Hermitian for general $\alpha \neq0$ so $FS^{(\alpha)}_{ji}\neq \bar{FS}^{(\alpha)}_{ij}$, where an overbar denotes the complex conjugation operation \footnote{This kind of structure naturally appears in the non-Hermitian generalisation of quantum mechanics \cite{Mostafazadeh}, where the left and the right eigenstates are not equal; see section \ref{NHQM} for further comments along this line.}.
Then, using the form of the generalised tensor structure \eqref{generaltensorstructure1}, we can obtain the generalised metric on this space given by the real and symmetric part, the explicit form of which can be written down as 
\begin{widetext}
    \begin{equation}
   \begin{split}
    g^{(\alpha)}_{ij}=\frac{1}{4}\mathcal{E}_{p}\Big[\cos{(2\alpha \phi)}\partial_{i}l_{\theta}\partial_{j}l_{\theta}\Big]+ \mathcal{E}_{p}\Big[ \cos{(2\alpha \phi)}\partial_{i}\phi\partial_{j}\phi\Big]-
    \frac{2}{(\alpha^2-1)}\Bigg(\mathcal{E}_{p}\Big[\cos{(2\alpha\phi)}\partial_{i}\phi\Big]\mathcal{E}_{p}\Big[\cos{(2\alpha\phi)}\partial_{j}\phi\Big]+\\
\mathcal{E}_{p}\Big[\cos{(2\alpha\phi)}\partial_{j}l_{\theta}\Big]\mathcal{E}_{p}\Big[\cos{(2\alpha\phi)}\partial_{i}l_{\theta}\Big]-
\mathcal{E}_{p}\Big[\sin{(2\alpha\phi)\partial_{i}}l_{\theta}\Big]\mathcal{E}_{p}\Big[\sin{(2\alpha\phi)\partial_{j}}l_{\theta}\Big]-4\mathcal{E}_{p}\Big[\sin{(2\alpha\phi)\partial_{i}}\phi\Big]\mathcal{E}_{p}\Big[\sin{(2\alpha\phi)\partial_{j}}\phi\Big]
    \Bigg)~.
\label{quantumalphametric1}
\end{split}
\end{equation}
\end{widetext}

This is one of the simplest generalised versions of the QMT, which we will call the $\alpha$- quantum-metric tensor (QAMT).
As an important cross-check, we can see that this reduces to the standard form of the QMT obtained in \cite{Facchi2010classical, Provost1980riemannian} for $\alpha=0$. 

In a similar way, we can also compute the general form of the antisymmetric, complex $2$-form, which we call the $\alpha$-Berry curvature from now on \footnote{For discussion of a proper form of the Berry curvature in the non-Hermitian case, see end of the section \ref{components}.}. The explicit form of the $\alpha$-Berry curvature is given by
\begin{widetext}
    \begin{equation}
\begin{split}
    \omega^{\alpha}_{ij}=\frac{i}{2(\alpha^2-1)}\Bigg((\alpha^2-1)\mathcal{E}_{p}\Big[\cos{(2\alpha\phi)}\Big(\partial_{i}l_{\theta}\partial_{j}\phi-\partial_{j}l_{\theta}\partial_{i}\phi\Big)\Big]-
\Big(\mathcal{E}_{p}\Big[\cos{(2\alpha\phi)}\partial_{j}l_{\theta}\Big]\mathcal{E}_{p}\Big[\cos{(2\alpha\phi})\partial_{i}\phi\Big]-\\\mathcal{E}_{p}\Big[\cos{(2\alpha\phi)}\partial_{i}l_{\theta}\Big]\mathcal{E}_{p}\Big[\cos{(2\alpha\phi)}\partial_{j}\phi\Big]+
\mathcal{E}_{p}\Big[\sin {(2\alpha\phi)}\partial_{i}l_{\theta}\Big]\mathcal{E}_{p}\Big[\sin{(2\alpha\phi)}\partial_{j}\phi\Big]-\mathcal{E}_{p}\Big[\sin {(2\alpha\phi)}\partial_{j}l_{\theta}\Big]\mathcal{E}_{p}\Big[\sin{(2\alpha\phi)}\partial_{i}\phi\Big]\Big)
    \Bigg)~.
\end{split}
\end{equation}
\end{widetext}
As can be checked, this reduces to the correct limit of \eqref{berrycuravture} for the particular case $\alpha=0$.
More importantly, due to the non-Hermitian structure of the $\alpha$-Fubini-Study tensor, we will have two new components, which are not present in the $\alpha=0$ case, which is the standard Hermitian tensor in the complex projective space. 
The real but antisymmetric $2$-form on this parameter manifold is given as 
\begin{widetext}
    \begin{equation}
\begin{split}
 \tilde{g}^{(\alpha)}_{ij}=\frac{1}{2(\alpha^2-1)}\Bigg((\alpha^2-1)\mathcal{E}_{p}\Big[\sin{(2\alpha\phi)}\Big(\partial_{j}l_{\theta}\partial_{i}\phi-\partial_{i}l_{\theta}\partial_{j}\phi\Big)\Big]-
\Big(\mathcal{E}_{p}\Big[\cos{(2\alpha\phi)}\partial_{j}l_{\theta}\Big]\mathcal{E}_{p}\Big[\sin{(2\alpha\phi})\partial_{i}\phi\Big]-\\\mathcal{E}_{p}\Big[\cos{(2\alpha\phi)}\partial_{i}l_{\theta}\Big]\mathcal{E}_{p}\Big[\sin{(2\alpha\phi)}\partial_{j}\phi\Big]+
\mathcal{E}_{p}\Big[\sin {(2\alpha\phi)}\partial_{j}l_{\theta}\Big]\mathcal{E}_{p}\Big[\cos{(2\alpha\phi)}\partial_{i}\phi\Big]-\mathcal{E}_{p}\Big[\sin{(2\alpha\phi)}\partial_{i}l_{\theta}\Big]\mathcal{E}_{p}\Big[\cos{(2\alpha\phi)}\partial_{j}\phi\Big]\Big)
    \Bigg)~.
\end{split}
\end{equation}
\end{widetext}

We can also obtain the form of the complex but symmetric $2$-form $\bar{\omega}^{(\alpha)}_{ij}$ as 
\begin{widetext}
\begin{equation}
\begin{split}
\tilde{\omega}^{(\alpha)}_{ij}=\frac{i}{4(\alpha^2-1)}\Bigg((\alpha^2-1)\mathcal{E}_{p}\Big[\sin{(2\alpha\phi)}\Big(\partial_{j}l_{\theta}\partial_{i}l_{\theta}+4\partial_{i}\phi\partial_{j}\phi\Big)\Big]+
\Big(\mathcal{E}_{p}\Big[\cos{(2\alpha\phi)}\partial_{j}l_{\theta}\Big]\mathcal{E}_{p}\Big[\sin{(2\alpha\phi})\partial_{i}l_{\theta}\Big]+\\
\mathcal{E}_{p}\Big[\cos{(2\alpha\phi)}\partial_{i}l_{\theta}\Big]\mathcal{E}_{p}\Big[\sin{(2\alpha\phi)}\partial_{j}l_{\theta}\Big]+
4\mathcal{E}_{p}\Big[\sin {(2\alpha\phi)}\partial_{j}\phi\Big]\mathcal{E}_{p}\Big[\cos{(2\alpha\phi)}\partial_{i}\phi\Big]+4\mathcal{E}_{p}\Big[\sin{(2\alpha\phi)}\partial_{i}\phi\Big]\mathcal{E}_{p}\Big[\cos{(2\alpha\phi)}\partial_{j}\phi\Big]\Big)
    \Bigg)~.
\end{split}
\end{equation}
\end{widetext}

As can be checked and as expected, all the components of the two tensors $\bar{\omega}^{(\alpha)}_{ij}$ and $\bar{g}^{(\alpha)}_{ij}$ vanish identically for the case $\alpha=0$, and these two tensors incorporate the novel features of our construction for the general $\alpha \neq 0$ cases. So this construction satisfies the two checks mentioned at the beginning of the section \ref{alphaconnectionsection} and we will check if the other two conditions mentioned there are also satisfied.

\subsubsection{Derivation from QFI}
In this subsection, we ask: if it is possible to obtain the form of this Fubini-Study structure, and subsequently the $\alpha$- quantum metric tensor from  (modified version of) quantum Fisher information, as was done in \cite{Facchi2010classical} for the case of $\alpha=0$. It turns out that, due to the lack of normalisation (for the two functions $l_{(\alpha)}, l_{(-\alpha)}$) it is not possible to obtain this simple form of generalised  Fubini-Study structure from the QFI \footnote{In fact, it is easy to see that in the presence of non-trivial phase $\phi$, the only way to get a structure like \eqref{alphametric} is to choose the polynomials to be $\alpha=0$, which reduces to the simple FS metric.}, however, this point will be our motivation for us to further generalise this construction, as we will outline in the next section. The key takeaway from this section is that it is necessary to go beyond the Hermitian FS tensor for generic quantum systems for a consistent $\pm\alpha$ representation of the tangent space.

\subsection{ Case-2: Asymmetric generalisation of the Fubini-Study metric}
Above, we have seen how the minimal modification of the Hermitian Fubini-Study structure might lead to the generalised and modified version of the QMT and the Berry curvature, which can be used to build metric connections; however, one of the problems of such simple modification is that it is not a consistent formulation for $\alpha\neq0$, due to the lack of normalisation. In this section, our primary motivation is to find somewhat more complicated modifications of the Fubini-Study tensor structure that preserve the normalisation, as well as have a well-defined $\alpha=0$ limit, such that all the relevant results for these cases are retained.   
To this end, we propose to work with the following one-parameter family of biorthogonal tensor structures, which are manifestly not Hermitian,
\begin{widetext}
  \begin{equation}
FS^{(\alpha)}_{ij}=\braket{\partial_{i}{l_{1(\alpha)}}|\partial_{j}l_{2(-\alpha)}}-(1-\alpha^2)\braket{\partial_{i}l_{1(\alpha)}|l_{2(-\alpha)}}\braket{l_{1(\alpha)}|\partial_{j}l_{2(-\alpha)}}~,
\label{generaltensorstructure2}
\end{equation}  
\end{widetext}
where the two functions $l_{1(\alpha)}(x;\theta)$ and $l_{2(\alpha)}(x;\theta)$ are asymmetric, phase shifted with respect to each other and are defined in terms of the polar decomposition of the wave function as
\begin{equation}
\begin{split}
      l_{1(\alpha)}(x;\theta)=\frac{P^{\frac{1-\alpha}{2}}}{1-\alpha}e^{i(1-\alpha)\phi}=\frac{\Psi^{1-\alpha}}{1-\alpha}~, ~~\text{and}~~\hspace{2mm}\\ l_{2(\alpha)}(x;\theta)=\frac{P^{\frac{1-\alpha}{2}}}{1-\alpha}e^{i(1+\alpha)\phi}~,  
\end{split}
\end{equation}
which shows that $ l_{1(\alpha)}$ is a polynomial of the wavefunction, and $ l_{2(\alpha)}$ is a phase shifted polynomial of the wavefunction.
Importantly, it is easy to check that the inner product of the two functions is normalised up to a constant; 
\begin{equation}
    \braket{l_{1(\alpha)}|l_{2(-\alpha)}}=\braket{l_{2(-\alpha)}|l_{1(\alpha)}}=\frac{1}{1-\alpha^2}~,
\end{equation}
which can be set to unity, a fact that is crucial in obtaining the tensor from the QFI for this set of inner products. As expected, neither $ l_{1(\alpha)}$, nor $ l_{2(\alpha)}$ is separately normalised to unity.
As was done in the original construction of Provost and Vallee in \cite{Provost1980riemannian}, we can obtain the $\alpha$-FS tensor from the expansion of the biorthogonal overlap function  of two nearby states, as 
\begin{widetext}
  \begin{equation}
\mathcal{D}^{\alpha}(l_{1(\alpha)},l_{2(-\alpha)}) =\braket{l_{1(\alpha)}(\theta+\delta\theta)-l_{1(\alpha)}(\theta)|l_{2(-\alpha)}(\theta+\delta\theta)-l_{2(-\alpha)}(\theta)}
\label{overlap2}
\end{equation}  
\end{widetext}
and demanding the invariance under a global phase transformation. Following this procedure, it is easy to check that, like the $\alpha=0$ case, the inner product $\braket{\partial_{i}l_{1(\alpha)}|\partial_{j}l_{2(-\alpha)}}$ is not invariant under a global phase transformation. Which can be made so after subtracting the second term of \eqref{generaltensorstructure2}.
This inner product structure can be considered as the bilinear form on the tangent vector spaces associated with two geometric data sets $l_{1(\alpha)}, l_{2(-\alpha)}$ pulled back to the parameter manifold, which acts as the common base manifold, which is linear in one and conjugate-linear in the other. Note that for the Hilbert space of square-integrable functions, the wavefunction $\Psi$ has to satisfy certain conditions for the polynomial combinations $\ket{l_{1(\alpha)}}$, and $\ket{l_{2(-\alpha)}}$ individually to be in the Hilbert space of square-integrable functions. Note that unlike the standard $\alpha$-divergence of information geometry, which satisfies the relation $D^{(\alpha)}(P||Q)=D^{(-\alpha)}(Q||P)$, for two normalised PDFs $P, Q$, here these overlap integrals satisfy a general conditions $\mathcal{D}^{(\alpha)}(\Psi||\Psi^{\prime})=(\mathcal{D}^{(-\alpha)}(\Psi^{\prime}||\Psi))^{**}$, where $^{**}$ represents, instead of just $\alpha\rightarrow -\alpha$, simultaneous transformation $1\rightarrow 2$ and taking complex conjugation for two normalised wavefunctions $\Psi$ and $\Psi^{\prime}$, a fact that will be reflected in the corresponding connections.

Also similar to the previous case, the form of the FS structure reduces to the standard Hermitian FS tensor on the complex projective space for $\alpha=0$; however, for general $\alpha$, this tensor is not manifestly Hermitian. It is to be noted that since we are using the biorthogonal inner product structure, it is more appropriate to think of the $\alpha$-FS tensor as a `distance' on the space of $\alpha$-density matrices introduced later. The biorthogonal structure and the discrimination between a state and the complex conjugate with respect to the inner product are strongly reminiscent of the non-Hermitian generalisation of quantum mechanics, for example, the seminal biorthogonal constructions in \cite{Brody2013biorthogonal, Curtright}. However, since our primary concern is the standard Hermitian paradigm, we will always explicitly construct Hermitian `observables', but as we will see, the inherent non-Hermitian nature of the $\alpha$-FS metric will leave imprints on the quantities like $\alpha$-quantum geometric tensor and the $\alpha$-Berry curvature in a non-trivial way.
As alluded to earlier, since our construction has a close relation with the non-Hermitian quantum mechanics, we expect that the two novel $2$-form constructed in here $\tilde{g}^{(\alpha)}_{ij}$ and $\tilde{\omega}^{(\alpha)}_{ij}$, should also provide non-vanishing contributions for these systems, and to the best of our knowledge has not been discussed in the literature extensively earlier.

\subsubsection{Components of the $\alpha$-FS tensor}
\label{components}
Let us first explicitly extract the form of the four components of the $\alpha$-FS tensor. The real and the symmetric tensor, which we call the $\alpha$-QMT, is given in terms of the probability distribution function and the phase of the position space of the wave function as 
\begin{widetext}
    \begin{equation}
    g^{(\alpha)}_{ij}=\frac{1}{4}\mathcal{E}_{p}\Big[\partial_{i}l_{\theta}\partial_{j}l_{\theta}\Big]+\frac{(1-\alpha)}{(1+\alpha)}\Bigg(\mathcal{E}_{p}\Big[\partial_{i}\phi\partial_{j}\phi\Big]-\mathcal{E}_{p}\Big[\partial_{i}\phi\Big]\mathcal{E}_{p}\Big[\partial_{j}\phi\Big]\Bigg)~,
\end{equation}
\end{widetext}
which, as expected, reduces to the standard form of QMT \eqref{FSmetric} in the $\alpha=0$ case. Notice that the `classical' contribution to the $\alpha$-QMT is independent of the parameter $\alpha$, similar to the classical Fisher-Rao metric; however, the non-trivial contribution of the phase factor does not, reflecting our normalisation choice. On the other hand, like the $\alpha=0$ case, the contribution from the phase is necessarily non-negative, and the associated classical distribution in the generic $\alpha\neq0$ in this case also provides the lower bound of the full quantum information metric \cite{Stout1}.  

Similarly, we obtain the form of the antisymmetric, purely imaginary part (which is the Berry curvature in the standard Hermitian case) by the following expression:
\begin{equation}
\omega^{(\alpha)}_{ij}=\frac{i}{2(\alpha+1)}\mathcal{E}_{p}\Big[\partial_{i}l_{\theta}\partial_{j}\phi-\partial_{i}\phi\partial_{j}l_{\theta} \Big].
\label{berrycuravturealpha}
\end{equation}
Following the same line, we obtain the form of the flipped rank-$2$ tensors, having the form;
\begin{equation}
    \tilde{g}^{(\alpha)}_{ij}=-\frac{i\alpha}{2(1+\alpha)}\mathcal{E}_{p}\Big[\partial_{i}l_{\theta}\partial_{j}\phi+\partial_{i}\phi\partial_{j}l_{\theta}\Big]~,
\end{equation}
which is the purely imaginary but symmetric part, and the real and antisymmetric part vanishes identically in this case $ \tilde{\omega}^{(\alpha)}_{ij}=0$. As it can be seen, for $\alpha=0$, the flipped tensors do not contribute to the FS tensor, which is Hermitian in that case. The above decomposition shows how we can build each tensor once we have the position space wavefunction for a pure quantum mechanical state, which is invariant under a global phase transformation, i.e., it is gauge-invariant under the simultaneous $U(1)$ operation $\ket{l_{1(\alpha)}}\rightarrow e^{i\beta}\ket{l_{1(\alpha)}}$ and $\ket{l_{2(-\alpha)}}\rightarrow e^{i\beta}\ket{l_{2(-\alpha)}}$~.

\subsubsection{Berry curvature}
Even though we have decomposed the non-Hermitian FS tensor, where the real, symmetric part can be ascribed to the $\alpha$-QMT \footnote{ See section \ref{optimisation1}, for the role of the real, symmetric and imaginary, symmetric parts of the $\alpha$-FS tensor.}, the significance of the rest of the parts is not entirely evident at the moment. In particular, it can be seen that the antisymmetric field strength of the $\alpha$-gauge fields, namely,  $\tilde{A}_{i}=i\braket{{l_{1(\alpha)}}|\partial_{j}l_{2(-\alpha)}}$ is, 
\begin{equation}
    \tilde{F}_{ij}=i\Big(\braket{\partial_{j}{l_{1(\alpha)}}|\partial_{i}l_{2(-\alpha)}}-\braket{\partial_{i}{l_{1(\alpha)}}|\partial_{j}l_{2(-\alpha)}}\Big)~,
\end{equation}
is actually equal to the sum of $\omega^{(\alpha)}_{ij}$ and $\tilde{\omega}^{(\alpha)}_{ij}$, i.e. the complex valued $\alpha$-Berry curvature in general non-Hermitian setting. Of course, since we are essentially working with an underlying Hermitian set-up, with a particular choice of normalisation, as we have seen above, $\tilde{\omega}^{(\alpha)}_{ij}$ vanishes identically, and the Berry curvature coincides with the antisymmetric, imaginary part of the full $\alpha$-FS tensor. However, as we discuss later in section \ref{LRberrycurvature}, this is one case of the general four types of Berry curvatures, which might arise in a biorthogonal situation, and in general, they are not equivalent \cite{Shen18, Kawabata}.

\subsubsection{Induced fluctuations }
\label{inducedfluctuation}
The physical significance of this $\alpha$- QMT and the $\alpha$- Berry curvature for a given quantum system can be illuminated by studying the case, where we assume the two independent states $\ket{l_{1(\alpha)}}$ and $\ket{l_{2(-\alpha)}}$ are constructed from a set of fixed, mutually biorthogonal states on the Hilbert space so that we can write 
\begin{equation}
     \ket{l_{1(\alpha)}}=e^{i s_{k}A^{k}_{1(\alpha)}} \ket{l^{0}_{1(\alpha)}}, ~~~\text{and} ~~  \ket{l_{2(-\alpha)}}=e^{i s_{k}A^{k}_{2(-\alpha)}} \ket{l^{0}_{2(-\alpha)}}~.
\end{equation}
Here we have assumed that the two sets of generators and the initial states are not related for general $\alpha\neq 0$, and are equal only for $\alpha=0$. However, the choice of normalisation for our case requires that at $s=0$, we have $\braket{l^{0}_{1(\alpha)}|l^{0}_{2(-\alpha)}}=1$, and also for the biorthogonal structure to remain valid we also have to impose, $\braket{l_{1(\alpha)}|l_{2(-\alpha)}}=1$ for all $s$. This means that the moralisation is maintained throughout the \textit{evolution}. The imposition of this condition requires certain constraints on the generators and the fixed initial states, which we will briefly discuss in the appendix \ref{Appendix1}. Then we can obtain the form of the $\alpha$-Fubini-Study tensor \eqref{generaltensorstructure2} for these states, for the case where the generators $A^{k}_{1(\alpha)}$ and $A^{k}_{2(-\alpha)}$ commute among themselves, which we have assumed for simplicity.
For a mutually commuting set of generators, such that, the set of operators $A^{i}_{1(\alpha)}$ commutes for different $i$ and similarly for $A^{i}_{2(-\alpha)}$.
With this assumption, we can readily obtain the following expression
\begin{widetext}
         \begin{equation}
(FS)_{ij}=\braket{l_{1(\alpha)}|(A^{i}_{1(\alpha)})^{\dagger}A^{j}_{2(-\alpha)}|l_{2(-\alpha)}}-\braket{l_{1(\alpha)}|(A^{i}_{1(\alpha)})^{\dagger}|l_{2(-\alpha)}}\braket{l_{1(\alpha)}|A^{j}_{2(-\alpha)}|l_{2(-\alpha)}}~,
    \label{alphaFSforcommutating}
    \end{equation}
\end{widetext}
from which we can also obtain the form of the metric tensor and the Berry curvature. Clearly, these are different orders of fluctuation moments for the generators in the states under study, the covariance of two operators under the non-Hermitian inner product, which is, in general, not zero. The difference from the standard $\alpha=0$ case can be gleaned from the fact that, for that case, the tensor structure \eqref{alphaFSforcommutating} reduces to $(FS)_{ij}=\braket{l^{0}|A^{i}A^{j}|l^{0}}-\braket{l^{0}|A^{i}|l^{0}}\braket{l^{0}|A^{j}|l^{0}}$ \cite{Heteneyi}, which being manifestly real, the Berry curvature is identically zero for any choice of commutating set of operators. However, due to the biorthogonal structure of our formalism, there is no guarantee that for a general set of operators, the $\alpha$-FS tensor is real for generic $\alpha\neq 0$, and there will be non-trivial $\alpha$- Berry-curvature on this parameter manifold, even when they commute among themselves.

\subsubsection{$\alpha$- Fubini-Study tensor from the $\alpha$-quantum Fisher information}
To obtain the form of the tensor structure, let us define for the pure quantum states $\Psi$, a generalised version of the standard density matrix $\rho$, which we call $\alpha$-density matrix given by \footnote{Of course these are not really density matrix of the system, which is still $\rho=\ket{\Psi}\bra{\Psi}$ for the pure states we are considering and satisfy all the usual properties.} 
\begin{equation}
    \rho^{(\alpha)}=\ket{l_{2(-\alpha)}}\bra{l_{1(\alpha)}}~.
\label{alphadensitymatrix}
\end{equation}
Importantly, note that this form of the $\alpha$-density matrices is not Hermitian 
$\rho^{(\alpha)}\neq(\rho^{(\alpha)})^{\dagger}$. However, due to the implementation of the normalisation condition, the traces of these operators are preserved; 
$\text{Tr}[\rho^{(\alpha)}]=\text{Tr}[(\rho^{(\alpha)})^{\dagger}]=\frac{1}{1-\alpha^2}$. Since the $\alpha$-density matrix is itself not Hermitian, the observable quantity can be constructed from these as $\rho^{(\alpha)}_{ob}=\frac{1}{2}\Big(\rho^{(\alpha)}+(\rho^{(\alpha)})^{\dagger}\Big)$~.

With respect to this $\alpha$-density matrix, we define the modified version of the $\alpha$-QFI, which, as expected, is not manifestly  Hermitian and depends explicitly on the parameter $\alpha$
\begin{equation}
\mathcal{F}_{g}:=(1-\alpha^2)^2\text{Tr}[\rho^{(\alpha)}\partial_{i}\rho^{(\alpha)}\partial_{j}\rho^{(\alpha)}]~.
\end{equation}
It is quite straightforward to show by using the form of the $\alpha$-density matrix and the normalisation conditions that we can arrive at the modified FS structure \eqref{generaltensorstructure2}.

\subsubsection{Expansion of the overlap integrals and the non-metric Connections}
\label{nmconnections1}
In this section, we will do a systematic expansion of the Provost-Vallee-like overlap integral for the non-Hermitian inner product \eqref{overlap2} and extract the metric and the connection coefficients in a similar way as is done in the classical formulation of information geometry \cite{Amari2000methods}. As was explained earlier, from the expansion of the overlap integral, we can identify the metric  \eqref{generaltensorstructure2} at the second order, while at the third order, we have, 
\begin{equation}
\tilde{\Gamma}_{ij,k}^{1(\alpha)}=\braket{\partial_{i}\partial_{j}{l_{1(\alpha)}}|\partial_{k}l_{2(-\alpha)}}, 
\label{alphaconnection}
\end{equation}
\text{and}
  \begin{equation}
\tilde{\Gamma}_{ij,k}^{2(-\alpha)}=\braket{\partial_{k}{l_{1(\alpha)}}|\partial_{i}\partial_{j}l_{2(-\alpha)}}~.
\label{alphamiconnection}
\end{equation}  

It can be easily checked that these objects are not tensors, but transform as connections under a generic coordinate transformation. Note that, since we have two types of functions in the formulation, which are not complex conjugate of each other, $\tilde{\Gamma}_{ij,k}^{1(\alpha)}$, and $\tilde{\Gamma}_{ij,k}^{2(-\alpha)}$, are not simply related to each other by $\alpha$ to $-\alpha$ transformation, indicated in our notations explicitly by the $1,2$ indices. Explicitly, 
$\tilde{\Gamma}_{ij,k}^{1(-\alpha)} \neq \tilde{\Gamma}_{ij,k}^{2(-\alpha)}$, rather we have to simultaneously perform  transformations $\alpha\rightarrow -\alpha$, $1\rightarrow2$, (or in reverse order) and take the complex conjugate to get the other connection starting from a particular one, 
\begin{equation}
    (\tilde{\Gamma}_{ij,k}^{2(-\alpha)})^{**}=(\tilde{\Gamma}_{ij,k}^{1(\alpha)})~,
\label{generaliseduality}
\end{equation}
where $**$ represents the three successive operations mentioned above and can be thought of as a generalisation of the classical duality. For $\alpha=0$, this is related to the connection obtained, for example, in \cite{Heteneyi}.

To understand the significance of these objects, we substitute the polar decomposition and obtain 
\begin{widetext}
\begin{equation}
\begin{split}
\tilde{\Gamma}_{ij,k}^{1(\alpha)}=\mathcal{E}_{p}\Big[\frac{1}{4}\Big(\underbrace{\partial_{i}\partial_{j}l_{\theta}+\frac{(1-\alpha)}{2}\partial_{i}l_{\theta}\partial_{j}l_{\theta}}_{\text{classical} \hspace{1mm}\alpha \hspace{1mm}\text{connection}}
\Big)\partial_{k}l_{\theta}-2i\frac{(1-\alpha)}{4}\partial_{(i}l_{\theta}\partial_{j)}\phi\partial_{k}l_{\theta}+\frac{(1-\alpha)^2}{(1+\alpha)}\partial_{(i}l_{\theta}\partial_{j)}\phi\partial_{k}\phi\\
-\frac{i}{2}\underbrace{\Big(\partial_{i}\partial_{j}\phi-i(1-\alpha)\partial_{i}\phi\partial_{j}\phi\Big)}_{\text{classical-quantum  }\hspace{1mm}\alpha \hspace{1mm}\text{connection}-1}\partial_{k}l_{\theta}+\frac{i(1-\alpha)}{2(1+\alpha)}\Big(\underbrace{\partial_{i}\partial_{j}l_{\theta}+\frac{(1-\alpha)}{2}\partial_{i}l_{\theta}\partial_{j}l_{\theta}}_{\text{classical-quantum }\hspace{1mm}\alpha \hspace{1mm}\text{connection}-2}\Big)\partial_{k}\phi+\frac{(1-\alpha)}{(1+\alpha)}\Big(\underbrace{\partial_{i}\partial_{j}\phi-i(1-\alpha)\partial_{i}\phi\partial_{j}\phi}_{\text{Phase contribution to}\hspace{1mm}\alpha \hspace{1mm}\text{connection}}\Big)\partial_{k}\phi\Big]~,
\end{split}
\end{equation}
\end{widetext}
and 
\begin{widetext}
    \begin{equation}
\begin{split}
\tilde{\Gamma}_{ij,k}^{2(-\alpha)}=\mathcal{E}_{p}\Big[\frac{1}{4}\Big(\underbrace{\partial_{i}\partial_{j}l_{\theta}+\frac{(1+\alpha)}{2}\partial_{i}l_{\theta}\partial_{j}l_{\theta}}_{\text{classical} \hspace{1mm}(-\alpha) \hspace{1mm}\text{connection}}
\Big)\partial_{k}l_{\theta}+2i\frac{(1-\alpha)}{4}\partial_{(i}l_{\theta}\partial_{j)}\phi\partial_{k}l_{\theta}+(1-\alpha)\partial_{(i}l_{\theta}\partial_{j)}\phi\partial_{k}\phi\\
+\frac{i(1-\alpha)}{2(1+\alpha)}\underbrace{\Big(\partial_{i}\partial_{j}\phi+i(1-\alpha)\partial_{i}\phi\partial_{j}\phi\Big)}_{\text{classical-quantum  }\hspace{1mm} (-\alpha) \hspace{1mm}\text{connection}-1}\partial_{k}l_{\theta}-\frac{i}{2}\Big(\underbrace{\partial_{i}\partial_{j}l_{\theta}+\frac{(1+\alpha)}{2}\partial_{i}l_{\theta}\partial_{j}l_{\theta}}_{\text{classical-quantum  }\hspace{1mm} (-\alpha) \hspace{1mm}\text{connection}-2}\Big)\partial_{k}\phi+\frac{(1-\alpha)}{(1+\alpha)}\Big(\underbrace{\partial_{i}\partial_{j}\phi+i(1-\alpha)\partial_{i}\phi\partial_{j}\phi}_{\text{Phase contribution}\hspace{1mm}(-\alpha) \hspace{1mm}\text{connection}}\Big)\partial_{k}\phi\Big]~,
\end{split}
\end{equation}
\end{widetext}
the classical part of which shows how the $\pm\alpha$ connections can be obtained in the trivial phase limit. From these expressions, we obtain the following conclusions: (1) as expected, this is symmetric in the first two coordinates $(i,j)$, (2) for systems having a trivial phase, this reduces to the $\alpha$ connection of classical information geometry, with a factor of $1/4$ coming similar to QMT \cite{Amari2000methods}, (3) since this is obtained from a non-Hermitian inner product, $\tilde{\Gamma}_{ij,k}^{(\alpha)}$ is not Hermitian and (4) most importantly, it is also not gauge-invariant, and does not transform like one under a simultaneous global redefinition $\ket{l_{1(\alpha)}}\rightarrow e^{i\beta}\ket{l_{1(\alpha)}}$ and $\ket{l_{2(-\alpha)}}\rightarrow e^{i\beta}\ket{l_{2(-\alpha)}}$, which leaves the norm invariant, hence this is not a physically meaningful connection on the parameter manifold. Similarly, expanding the eq. \eqref{overlap2}, we can obtain the modified versions of the $-\alpha$ connection, modified by the non-trivial phase of the system.

Another important aspect to note is that, with respect to the bare  symmetric tensor $FS^{sy(\alpha)}_{ij}=\frac{1}{2}\Big(FS^{(\alpha)}_{ij}+FS^{(\alpha)}_{ji}\Big)$ (not the gauge-invariant version) and the two connections $\tilde{\Gamma}_{ij,k}^{(\pm\alpha)}$, they satisfy, 
\begin{equation}
\partial_{k}FS^{sy(\alpha)}_{ij}=\frac{1}{2}\Big(\tilde{\Gamma}_{ik,j}^{1(\alpha)}+\tilde{\Gamma}_{jk,i}^{2(-\alpha)}+\tilde{\Gamma}_{jk,i}^{1(\alpha)}+\tilde{\Gamma}_{ik,j}^{2(-\alpha)}\Big)~,
\end{equation}
which is reminiscent of the famous $\pm\alpha$ duality in classical information geometry, however, we emphasise that it is not meaningful to interpret this as a quantum generalisation of such duality, as all the quantities appearing in the expressions are not gauge-invariant, hence do not carry any physical meaning. To obtain a proper $\pm\alpha$- like duality in the quantum regime, we have to first remove the gauge redundancy in the corresponding connections, a task we perform in the next subsection.

\subsubsection{Gauge-invariant connections}
It is straightforward to show that, even though neither connections \eqref{alphaconnection} and \eqref{alphamiconnection} are invariant under gauge transformations $\ket{l_{1(\alpha)}}\rightarrow e^{i\beta}\ket{l_{1(\alpha)}}$ and $\ket{l_{2(-\alpha)}}\rightarrow e^{i\beta}\ket{l_{2(-\alpha)}}$, the following combinations are 
\begin{widetext}
\begin{equation}
\begin{split}
\Gamma_{ij,k}^{1(\alpha)}=\braket{\partial_{i}\partial_{j}{l_{1(\alpha)}}|\partial_{k}l_{2(-\alpha)}} -(1-\alpha^2)\Big(\braket{\partial_{i}\partial_{j}{l_{1(\alpha)}}|l_{2(-\alpha)}}\braket{{l_{1(\alpha)}}|\partial_{k}l_{2(-\alpha)}}+2\braket{\partial_{(i}{l_{1(\alpha)}}|l_{2(-\alpha)}}\braket{\partial_{j)}{l_{1(\alpha)}}|\partial_{k}l_{2(-\alpha)}}\Big)\\
+2(1-\alpha^2)^2\braket{\partial_{i}{l_{1(\alpha)}}|l_{2(-\alpha)}}\braket{\partial_{j}{l_{1(\alpha)}}|l_{2(-\alpha)}}\braket{{l_{1(\alpha)}}|\partial_{k}l_{2(-\alpha)}}~, 
\label{alphaconnectiongauged}   
\end{split}
\end{equation}
\end{widetext}
and 
\begin{widetext}
    \begin{equation}
\begin{split}
\Gamma_{ij,k}^{2(-\alpha)}=\braket{\partial_{k}{l_{1(\alpha)}}|\partial_{i}\partial_{j}l_{2(-\alpha)}}-(1-\alpha^2)\Big(\braket{{\partial_{k}l_{1(\alpha)}}|l_{2(-\alpha)}}\braket{{l_{1(\alpha)}}|\partial_{i}\partial_{j}l_{2(-\alpha)}}+2\braket{{l_{1(\alpha)}}|\partial_{(i}l_{2(-\alpha)}}\braket{{\partial_{k}l_{1(\alpha)}}|\partial_{j)}l_{2(-\alpha)}}\Big)\\
+2(1-\alpha^2)^2\braket{\partial_{k}{l_{1(\alpha)}}|l_{2(-\alpha)}}\braket{{l_{1(\alpha)}}|\partial_{i}l_{2(-\alpha)}}\braket{{l_{1(\alpha)}}|\partial_{j}l_{2(-\alpha)}}~.
\label{alphamconnection}
\end{split}
\end{equation}
\end{widetext}

The $\alpha=0$ case of these connections was obtained in \cite{Heteneyi}, from a gauge-invariant generating function (with an overall trivial phase factor); however, here we have chosen to first have the bare connection and then make it gauge-invariant. Importantly, the gauge-invariant versions of the tensors $\ket{\partial_{i}\partial_{j}l_{2(-\alpha)}}$ or $\bra{\partial_{i}\partial_{j}l_{1(\alpha)}}$ are not symmetric in the two indices, and the difference measures the curvature of the parameter manifold with respect to the connections induced by the non-Hermitian inner product. Then, it can be checked that, with respect to the gauge-invariant $\alpha$-FS tensor and the two connections formally, we indeed have, 
\begin{equation}
\partial_{k}FS^{sy(\alpha)}_{ij}=\frac{1}{2}\Big(\Gamma_{ik,j}^{1(\alpha)}+\Gamma_{jk,i}^{2(-\alpha)}+\Gamma_{jk,i}^{1(\alpha)}+\Gamma_{ik,j}^{2(-\alpha)}\Big)~,
\label{pmalphaduality}
\end{equation}
which do resemble $\pm\alpha$ duality (with proper symmetrisation, as the FS tensor is not symmetric)  for physically meaningful gauge-invariant tensor structures on the parameter manifold; however, since we are dealing with generic non-Hermitian tensors, whether the apparent duality \eqref{pmalphaduality} can be interpreted as the existence of affine coordinates like the classical case remains to be seen. It is also possible to obtain the symmetric combinations of the $1(\alpha), 2(-\alpha)$ connections as symmetric combinations of $\mathcal{F}_{\Gamma^{1,2}}:=(1-\alpha^2)^4\text{Tr}\Big[(1-\alpha^2)^2\text{Tr}[\rho^{(\alpha)}\partial_{(i}\rho^{(\alpha)}]\rho^{(\alpha)}\partial_{j)}\rho^{(\alpha)}\partial_{k}\rho^{(\alpha)}\rho^{(\alpha)}\Big]$, which is the physically meaningful symmetric and gauge-invariant quantity in this context.

\subsubsection{Decomposition of the $\alpha$-connections}
Since both the $\pm\alpha$-connections obtained in the earlier sections are already symmetric in the first two indices, in this section we will focus on the real and imaginary part of these gauge-invariant connections and their relation with the metric connections obtained in section \ref{metricocnnection}, which is manifestly real by construction. The expression of $\Gamma_{ij,k}^{1(\alpha)}$ in terms of the two real functions $P$ and $\phi$, is of the form

\begin{widetext}
    \begin{equation}
\begin{split}
\Gamma_{ij,k}^{1(\alpha)}=\tilde{\Gamma}_{ij,k}^{1(\alpha)}
-(1-\alpha^2)\Bigg(\frac{i}{2(1+\alpha)^2}\mathcal{E}_{p}\Big[\partial_{i}\partial_{j}l_{\theta}+\frac{(1-\alpha)}{2}\partial_{i}l_{\theta}\partial_{j}l_{\theta}
\Big]\mathcal{E}_{p}\Big[\partial_{k}\phi\Big]+\frac{1-\alpha}{(1+\alpha)^2}\mathcal{E}_{p}\Big[\partial_{(i}l_{\theta}\partial_{j)}\phi\Big]\mathcal{E}_{p}\Big[\partial_{k}\phi\Big]+\\
\frac{1}{(1+\alpha)^2}\mathcal{E}_{p}\Big[\partial_{i}\partial_{j}\phi-i(1-\alpha)\partial_{i}\phi\partial_{j}\phi\Big]\mathcal{E}_{p}\Big[\partial_{k}\phi\Big]-\frac{i}{2(1+\alpha)}\mathcal{E}_{p}\Big[\partial_{(i}\phi\Big]\mathcal{E}_{p}\Big[\partial_{j)}l_{\theta}\partial_{k}l_{\theta}+\frac{4(1-\alpha)}{1+\alpha}\partial_{j)}\phi\partial_{k}\phi\Big]\\
-\frac{1}{(1+\alpha)}\mathcal{E}_{p}\Big[\partial_{(i}\phi\Big]\mathcal{E}_{p}\Big[\partial_{j)}\phi\partial_{k}l_{\theta}-\frac{1-\alpha}{1+\alpha}\partial_{j)}l_{\theta}\partial_{k}\phi\Big]
\Bigg)-2i\frac{(1-\alpha)^2}{1+\alpha}\mathcal{E}_{p}\Big[\partial_{i}\phi\Big]\mathcal{E}_{p}\Big[\partial_{i}\phi\Big]\mathcal{E}_{p}\Big[\partial_{k}\phi\Big].
\end{split}
\end{equation}
\end{widetext}
Similarly, we can obtain the following, 
\begin{widetext}
    \begin{equation}
\begin{split}
\Gamma_{ij,k}^{2(-\alpha)}=\tilde{\Gamma}_{ij,k}^{2(-\alpha)}
-(1-\alpha^2)\Bigg(-\frac{i}{2(1-\alpha^2)}\mathcal{E}_{p}\Big[\partial_{i}\partial_{j}l_{\theta}+\frac{(1+\alpha)}{2}\partial_{i}l_{\theta}\partial_{j}l_{\theta}
\Big]\mathcal{E}_{p}\Big[\partial_{k}\phi\Big]+\frac{1}{(1+\alpha)}\mathcal{E}_{p}\Big[\partial_{(i}l_{\theta}\partial_{j)}\phi\Big]\mathcal{E}_{p}\Big[\partial_{k}\phi\Big]+\\
\frac{1}{(1+\alpha)^2}\mathcal{E}_{p}\Big[\partial_{i}\partial_{j}\phi+i(1-\alpha)\partial_{i}\phi\partial_{j}\phi\Big]\mathcal{E}_{p}\Big[\partial_{k}\phi\Big]+\frac{i}{2(1+\alpha)}\mathcal{E}_{p}\Big[\partial_{(i}\phi\Big]\mathcal{E}_{p}\Big[\partial_{j)}l_{\theta}\partial_{k}l_{\theta}+\frac{4(1-\alpha)}{1+\alpha}\partial_{j)}\phi\partial_{k}\phi\Big]\\
-\frac{1-\alpha}{(1+\alpha)^2}\mathcal{E}_{p}\Big[\partial_{(i}\phi\Big]\mathcal{E}_{p}\Big[\partial_{j)}\phi\partial_{k}l_{\theta}-\frac{1+\alpha}{1-\alpha}\partial_{j)}l_{\theta}\partial_{k}\phi\Big]
\Bigg)+2i\frac{(1-\alpha)^2}{1+\alpha}\mathcal{E}_{p}\Big[\partial_{i}\phi\Big]\mathcal{E}_{p}\Big[\partial_{i}\phi\Big]\mathcal{E}_{p}\Big[\partial_{k}\phi\Big]~.
\end{split}
\end{equation}
\end{widetext}

A few observations are in order: (1) first note that for the $\alpha=0$ case, the real part of both $\Gamma_{ij,k}^{1(\alpha)}$ and $\Gamma_{ij,k}^{2(-\alpha)}$ reduces to the metric connection obtained in section \ref{symmetricompatibleconnection}, which can be explicitly seen from eq. \eqref{metricicnnevtionPphi}. Since this $\Gamma^{(c)}_{ij,k}$ is built solely from the real symmetric metric, it cannot capture the specific phase contribution in the $\alpha\neq 0$ connections. (2) Similarly, the real parts of the $\pm$ connections satisfy the following duality 
\begin{equation}
\text{Re}[\Gamma_{ij,k}^{1(\alpha)}+\Gamma_{ij,k}^{2(-\alpha)}]= 2\Gamma^{(c)}_{ij,k}~,
\end{equation}
which is indeed similar to the classical case, here generalised to include the phase contribution in a gauge-invariant way for a pure state, and the sum of the (real parts) of the two-metric connections gives the metric connection with respect to the QMT. This shows that, for Hermitian observables properly built from the connections, it is possible to get $\pm$-duality with respect to the two connections and the real part of the FS tensor, including the contribution for a non-trivial phase.

\subsection{Optimisation problem with quantum natural gradient for $\alpha$-FS tensor}
\label{optimisation1}
The natural gradient direction in any statistical manifold moves in the steepest descent direction according to the inherent geometry of the manifold, encoded in the FR metric and the two dual connections \cite{Amari2000methods}, and has significant advantages over other standard gradient descents \cite{Amari1998}. In a similar motivation, the geometry of the space of quantum states was used in \cite{Stokes, Yamamoto} to provide the concept of a natural quantum gradient for variational algorithms for quantum systems. In this interpretation, the QMT (of the standard Hermitian QGT) chooses a direction that minimises the loss function for the inner product defined with respect to the QMT. 

We will first briefly describe the natural gradient optimisation with respect to the QMT in the standard Hermitian FS tensor. To this end, we consider a set of states $\Psi_{\theta}$, parametrised by a set of parameters $\theta$. As was considered, for example, in section \ref{inducedfluctuation}, this set of states can be thought of as obtained from a set of unitary operators, parametrised by $\{\theta\}$, acting on a fixed state $\ket{0}$ on the Hilbert space. Then for a Hermitian operator $H=H^{\dagger}$ in the Hilbert space, we want to optimise $L(\theta)=\braket{\Psi_{\theta}|H|\Psi_{\theta}}$, the real-valued cost, subject to the step size defined by the QMT, $g_{ij}^{FS}$. Then the local solution of the optimisation problem for a small variation of the parameter $\delta\theta^{i}$ is,
\begin{equation}
   g_{ij}^{FS}\delta\theta^{j}=-\eta\partial_{i}L(\theta)~,  \end{equation}
such that the optimal direction on the parameter manifold is 
\begin{equation}
    \theta_{t+1}^{i}=\theta^{i}_{t}-\eta g^{(FS)ij}\partial_{j}L(\theta)~,
\end{equation}
provided that the metric is invertible.  This shows that for optimisation, the point $\theta^{i}_{t}$ must move in the \textit{opposite} direction to the natural gradient of the loss function with respect to $ g_{ij}^{FS}$. This also provides an interpretation of the metric as an indicator of the natural `flow' on the manifold.
Note that for states generated by Hermitian operators, $\partial_{i}L(\theta)$ is always real. 

Now, we will interpret the symmetric part of the $\alpha$-FS tensor (both the real and the imaginary parts) as an indicator of the natural flow directions for a similar optimisation problem, but for generic non-Hermitian operators. Let us consider two sets of parametrised states $\bra{l_{1(\alpha)}(\theta)}$ and $\ket{l_{2(-\alpha)}(\theta)}$, which can be thought to be generated by two sets of generators, which need not be Hermitian. Then we consider a modified loss function of the following form: $\tilde{L}(\theta)=\braket{l_{1(\alpha)}(\theta)|A|l_{2(-\alpha)}(\theta)}$, which, in general, is a complex-valued function, and we denote the real and the imaginary parts as $\tilde{L}_R$,  $\tilde{L}_{I}$ respectively. This requires us to consider two \textit{simultaneousness} optimisation problems, one for $\tilde{L}^R$ and the other for $\tilde{L}^{I}$. Then the solution to the two optimisation problems can be similarly written down as 
\begin{equation}
   g_{ij}^{\alpha}\delta\theta^{j}=-\eta_{R}\partial_{i}\tilde{L}_{R}(\theta) ~~\text{and}\hspace{2mm} ~\tilde{g}_{ij}^{\alpha}\delta\theta^{j}=-\eta_{I}\partial_{i}\tilde{L}_{I}(\theta)~,
\label{nonhermitianoptimisation}
\end{equation}
where we have used the purely real, symmetric and purely imaginary, symmetric decomposition of the $\alpha$-FS tensor. This illustrates the role of the real and the imaginary component of the full non-Hermitian metric: the real part dictates the flow of the real part of the generic non-Hermitian `observable' under measurement, and the imaginary part governs the optimised flow of the imaginary part of the loss function. However, note that, for the real parameter space, the set of equations described by \eqref{nonhermitianoptimisation}, may not, and in general, will not be compatible with each other. As a result, the natural flow for the real and the imaginary part of the loss function will not, in general, be compatible - a fact to be considered when optimising such systems. Note also that, if we consider Hermitian observable, which is the case in the present context, even in a biorthogonal system, that will `see' only the real and symmetric part of the full non-Hermitian FS tensor, hence the proper interpretation requires the specification of the observable under study, a point we will return when discussing the role of different decomposition of QGT for manifestly non-Hermitian systems in section \ref{NHQM}. 

\section{Quantum- geometric tensor in the non-Hermitian quantum mechanics}
\label{NHQM}
The natural emergence of non-Hermitian structures in the biorthogonal formalism motivates us to briefly consider the geometry of quantum states governed by a non-Hermitian Hamiltonian \cite{Cui, BrodyIGNh, Zhangquantum, Zhuband, Solnyshkovquantum, Chen, Cuerda, Hu1, Hu2, Alon, Lu24} and even though they agree in the Hermitian limit, the definitions used for QMT and Berry curvatures are often different. The primary source of disagreement seems to be two-fold. 
First, generalizations of the QGT from the Hermitian case do not necessarily yield a real-valued metric when symmetrizing over parameter indices. This leads to multiple inequivalent options~\cite{Chen}---namely, to take either the \textit{real part}, the \textit{symmetric part}, or the \textit{real and symmetric part} of the QGT as QMT, each of which has appeared in the literature.
On the other hand, since the left and right states are not complex conjugates of each other, it is often interesting to define and study tensors constructed only from the right \textit{ or} only from the left states, and in general they do not capture the same information about a physical system \cite{Chen, Hu1, Hu2} \footnote{This only refers to a very specific set of works on various aspects of non-Hermitian systems, for a more complete overview and references, the reader is referred to \cite{Mostafazadeh, Ashida}}. In our opinion, the way to define a notion of inner product and the corresponding metric will depend on the context in which it is used and, in particular, on the way the states are normalised, as the four different ways of defining metric tensor $(FS)^{LR}$ or$(FS)^{RL}$ or $(FS)^{LL}$, or $(FS)^{RR}$  are not equivalent. Here we provide a systematic classification of all such metrics (and other tensors) based on two first principles $(1)$ the nature of the normalisation condition and inner product implemented and $(2)$ the `observable' under study, i.e., if we represent the optimal evolution of an observable on the parameter manifold, then depending on the nature of the observable (Hermitian or not), it will \textit{see} different parts of the full non-Hermitian FS tensor that we introduce later.  These two conditions will provide a systematic characterisation of the four components of the non-Hermitian FS tensor and in which context they are to be used. Throughout the rest of the paper we will use the standard notation of using $\ket{\Psi_{R}}$ as the right state of a given non-Hermitian Hamiltonian, which can be written in terms of the complete, but in general non-orthogonal to each other, right eigenstates of the Hamiltonian. The Hermitian adjoint of this state will be denoted as $\bra{\Psi_{R}}$, which again can be expressed as a linear combination of the Hermitian adjoint of the right eigenstates, which are not the left eigenstates of the Hamiltonian. On the other hand we will denote the left state as $\bra{\Psi_{L}}$, can be written as a combination of the left eigenstates of the Hamiltonian and the  biorthogonal inner product between a right and a left state will be denoted as $\braket{\Psi^{L}|\Psi^{R}}$.

\subsection{LR-Fubini-Study tensor}
\label{LRFStensorsection}
The most natural way to arrive at the QGT for this case, and the corresponding decomposition in QMT and the Berry curvature, is to start from the non-Hermitian inner product on the base manifold, equipped with the local coordinate chart defined by the parameters of the system, where the two domains of the left and right states are pulled back to, as
\begin{equation}
FS_{ij}^{LR}=\braket{\partial_{i}\Psi^{L}|\partial_{j}\Psi^{R}}-\braket{\partial_{i}\Psi^{L}|\Psi^{R}}\braket{\Psi^{L}|\partial_{j}\Psi^{R}}~, 
\label{FSleftright}
\end{equation}
which is explicitly not Hermitian due to the bilinear combination of left and right states. Note that here we are using the Hermiticity condition in the standard way; the tensor $FS_{ij}^{LR}$, however is `Hermitian' with respect to the `bilinear conjugate transpose' operation, which essentially is the origin of the $1-2$ duality apart from the standard $\pm \alpha$ one in the above construction in \eqref{generaliseduality}. 
Here, the last term included ensures the gauge-invariance under the transformation $\ket{\Psi^{L}} \rightarrow e^{i\beta_{L}}\ket{\Psi^{L}} $ and $\ket{\Psi^{R}} \rightarrow e^{i\beta_{R}}\ket{\Psi^{R}} $, provided that they satisfy certain conditions, as we explain later. This can be obtained from a non-Hermitian overlap like Provost-Vallee, suitably generalised in this case as $\mathcal{D}^{LR}(\theta, \theta^{\prime})=\braket{\Psi^{L}(\theta+\delta\theta)-\Psi^{L}(\theta)|\Psi^{R}(\theta+\delta\theta)-\Psi^{R}(\theta)}$, and demanding the invariance under the simultaneous gauge transformations of the left and right states. At this point, we want to emphasise that it is extremely important to note that this exercise is only meaningful when the normalisation of the states is such that
\begin{equation}
    \braket{\Psi^{L}|\Psi^{R}}=1,
    \label{leftrightnorm}
\end{equation}
which is explicitly used in the Provost-Vallee-like approach, a fact also stressed in \cite{Chen, Hu1}. This condition is assumed to be valid at \textit{every} instance of evolution on the parameter manifold, a condition, which is not guaranteed to be satisfied under evolution by the non-Hermitian generator, since the resulting state does not belong to the unitary orbit of $\bra{\Psi^{L}_{0}}$ and $\ket{\Psi^{R}_{0}}$. However, if we use the associated left states of a given right states, then this is guaranteed once we fix the normalisation. Another important point to note is that this normalisation is invariant under the local gauge transformation of the form $\Psi^{L} \rightarrow e^{i\beta_{L}}\Psi^{L}$ and $\Psi^{R} \rightarrow e^{i\beta_{R}}\Psi^{R}$, where one now does not imply other; hence, they have to satisfy the condition $\beta_{L}=\beta_{R}\in \mathbb{R}$, for the norm to remain invariant under such a transformation.   It is also straightforward to show that such a $2$-form \eqref{FSleftright} can be obtained from a QFI-like quantity for the non-Hermitian `density matrix' $\rho_{RL}=\ket{\Psi^{R}}\bra{\Psi^{L}}$, and using the trace condition along with the adopted normalisation. 

Then the `metric' on the base manifold that measures the distance between nearby left states and nearby right states, as was shown in the overlap above, can be obtained as the real \textit{and} symmetric part of the $FS^{LR}$ tensor as 
\begin{equation}
    g_{ij}^{LR}=\frac{1}{4}\Big[(FS_{ij}^{LR})+\bar{(FS_{ij}^{LR})}+(FS_{ji}^{LR})+\bar{(FS_{ji}^{LR})}\Big]~.
\label{metricleftright}
\end{equation} 
This essentially measures the \textit{response} of the pulled-back Hilbert-space inner product when the left state is changed in the $i$th direction and the right state is in the $j$th direction on the parameter manifold (though as we will see later, this also receives contribution from the imaginary and symmetric part also). On the other hand, the purely imaginary \textit{and} antisymmetric 2-form constructed from the $FS^{LR}$ is 
\begin{equation}
     \omega_{ij}^{LR}=\frac{1}{4}\Big[\big((FS_{ij}^{LR})-\bar{(FS_{ij}^{LR})}\big)-\big((FS_{ji}^{LR})-\bar{(FS_{ji}^{LR})}\big)\Big]~.
\label{berryleftright}
\end{equation}
Perhaps the most important outcome of decomposing the FS tensor in this is that this leaves two independent tensors: those are the real \textit{but} antisymmetric (we call it the flipped (part of the) QMT) part of $FS^{LR}$ and the purely imaginary \textit{but} symmetric part of $FS^{LR}$ (we call it the flipped (part of the)  Berry curvature). They can be written down explicitly as 
\begin{equation}
    \tilde{g}_{ij}^{LR}=\frac{1}{4}\Big[\big((FS_{ij}^{LR})-\bar{(FS_{ij}^{LR})}\big)+\big((FS_{ji}^{LR})-\bar{(FS_{ji}^{LR})}\big)\Big]~,
\end{equation}
and 
\begin{equation}
    \tilde{\omega}_{ij}^{LR}=\frac{1}{4}\Big[\big((FS_{ij}^{LR})+\bar{(FS_{ij}^{LR})}\big)-\big((FS_{ji}^{LR})+\bar{(FS_{ji}^{LR})}\big)\Big]~.
\end{equation}
In our opinion, these four tensors provide a systematic and clear classification of the contributions from the non-Hermitian tensor structures, which signifies different contributions to the state-space geometry. However, at this point it is not evident how the two flipped tensor contributes to the geometry, the role of which will be clarified later when we consider the natural optimisation problem on this (base) parameter manifold.

Finally, even though we have interpreted the LR-FS tensor on the space of states, it will also have an equivalent interpretation on the space of LR-density matrices, $\rho^{LR}(\theta)=\ket{\Psi^{R}}\bra{\Psi^{L}}$, which will lead to a natural extension for the mixed states. For pure states then, we will have, $\rho^{LR} = |\Psi_R\rangle\langle\Psi_L|$, satisfying $\rho^2 = \rho$, $\mathrm{Tr}^\# \rho = 1$, and $\rho^\# = \rho$, where the $\#$ indicates operations respecting the biorthogonal structure. At this point, rather speculatively, let us define the bi-symmetric logarithmic derivative (BSLD) $L^{LR}_{\theta}$ at least formally, via 
\begin{equation}
   \partial_{i} \rho^{LR}= \frac{1}{2}(L^{LR}_{\theta}\rho^{LR}_{\theta} + \rho^{LR}_{\theta} L^{LR}_{\theta}), \quad \text{with } L^{(LR)\#}_{\theta} = L^{LR}_{\theta}. 
\end{equation}
This does mirror the Hermitian condition $L_\theta = L^\dagger_\theta$, replacing the adjoint with the biorthogonal $\#$-adjoint. The corresponding biorthogonal quantum Fisher information is $\mathcal{F}^{LR} = \mathrm{Tr}^\#\Big[\rho^{LR}_{\theta} \partial_{i}L^{LR}_{\theta}\partial_{j}L^{LR}_{\theta}\Big]$, and is exactly equivalent to the (four times) the LR-FS tensor \eqref{FSleftright} for properly normalised left-right states. However, how far the analogy with the standard Hermitian case can be extended and what implications they have for parameter estimations remains a topic of further study.

\subsubsection{Non-Hermitian extension of Berry curvature}
\label{LRberrycurvature}
To understand the role of each component, let us first note that, for the complex-valued Berry connection in this case $\tilde{A}_{i}=i\braket{\Psi^{L}|\partial_{j}\Psi^{R}}$, the curvature of this complex $2$-form is 
\begin{equation}
    \tilde{F}^{LR}=\partial_{i}\tilde{A}_{j}-\partial_{j}\tilde{A}_{i}= i\Big(\braket{\partial_{i}\Psi^{L}|\partial_{j}\Psi^{R}}-\braket{\partial_{j}\Psi^{L}|\partial_{i}\Psi^{R}}\Big)~.
\end{equation}
This is essentially the antisymmetric part of $FS_{ij}^{LR}$, i.e. the sum of  $\omega_{ij}^{LR}$ and $\tilde{\omega}_{ij}^{LR}$, is the analogue of the Berry curvature in the non-Hermitian case. As is evident, once we have chosen a particular inner product (and normalisation with respect to that inner product, which also dictates the $U(1)$ gauge transformation), the form of the Berry curvature is well-defined and unique. Of course, we could also define the inner product as $\braket{\Psi^{R}|\Psi^{L}}=1$, depending on the contex  for a given set of left and right set of states, then it is obvious that the form of the metric, Berry curvature will also be changed, and can be obtained from the right-left FS tensor of the form $FS_{ij}^{RL}=\braket{\partial_{i}\Psi^{R}|\partial_{j}\Psi^{L}}-\braket{\partial_{i}\Psi^{R}|\Psi^{L}}\braket{\Psi^{R}|\partial_{j}\Psi^{L}}$, which in turn is the result of the expansion of the overlap integral of the form $\braket{\Psi^{R}(\theta+\delta\theta)-\Psi^{R}(\theta)|\Psi^{L}(\theta+\delta\theta)-\Psi^{L}(\theta)}$, and in general they will not be equivalent. These two cases would then give rise to what was called the $B^{LR}$ and $B^{RL}$ Berry curvature in \cite{Shen18} (see also \cite{Kawabata, Fan, Hu1}). It is important to mention that if we consider a pair of associated right and left states, then, even though the biorthogonal inner products $\braket{\Psi^{R}|\Psi^{L}}$ and  $\braket{\Psi^{L}|\Psi^{R}}$, refer to mapping from completely different vector spaces, at least for finite-dimensional Hilbert spaces, one fix normalisation do imply the other. However, for a generic biorthogonal combination it is important to fix the normalisation in the two cases separately, and it will not, in general be consistent with one another.



\subsubsection{Optimisation problem and the LR (RL) QMT}
\label{optimisation2}
As was discussed in section \ref{optimisation1}, the quantum natural gradient descent techniques uses the natural metric on the quantum state space to optimise a given cost, which in the quantum variational eigensolver problem is the expectation value of the (Hermitian) Hamiltonian, and the zero variance conditions are employed to extract the eigenstates and the eigenvalues. However, in a non-Hermitian setting, this cost, in general, will not be a real-valued function of the parameters, and such a cost can be decomposed as $\mathcal{L}^{LR}=\braket{\Psi^{L}(\theta)|A|\Psi^{R}(\theta)}$, for a non-Hermitian operator $A$. Here we have assumed that the left and right set of states is generated by a set of left and right generators, parametrised by $\{\theta_{i}\}$, from a fixed left and right reference state $\bra{\Psi^{L}_{0}}$ and $\ket{\Psi^{R}_{0}}$. Then, in the quantum natural gradient descent approach, the flow of a point $\theta_{i}$ is opposite to the gradient of the cost at that point, with respect to the underlying geometry of the curved manifold, which in the standard Hermitian case is the QMT.

In the non-hermitian case, on the other hand, to optimise the cost $\mathcal{L}^{LR}=\braket{\Psi^{L}(\theta)|A|\Psi^{R}(\theta)}$, defined for a pair of left and right states, it is most convenient to generalise the Hermitian quantum natural gradient to the two simultaneous optimisation problem for the real-valued costs $R(\mathcal{L}^{LR})$ and $I(\mathcal{L}^{LR})$, the real and the imaginary part of the full cost. Then this optimisation  problem with the imposed constraint determined by the step size, controlled by the symmetric part of $FS_{ij}^{LR}$ will provide two mutually incompatible solutions for the real parameter, each for real and the imaginary part of the cost, where the individual gradient is with respect to the real and the imaginary component of the symmetric part of the FS tensor, 
\begin{equation}
   g_{ij}^{LR}\delta\theta^{j}=-\eta^{LR}_{R}\partial_{i}R\Big(\mathcal{L}^{LR}(\theta)\Big)~, ~~\text{and}~~\hspace{2mm} \tilde{g}_{ij}^{LR}\delta\theta^{j}=-\eta^{LR}_{I}\partial_{i}I\Big(\mathcal{L}^{LR}(\theta)\Big)~.
\label{LRoptimisation}
\end{equation}
This illustrates the role of the individual components of the non-Hermitian QMT and, importantly, implies that the `natural flow' direction of each component may be different from each other. Note that, even though the symmetric part of $FS_{ij}^{LR}$, which is equal to $g_{ij}^{LR}+\tilde{g}_{ij}^{LR}$ governs the dynamics of the variational problem, it will not be appropriate to call it a `metric' in the standard use of this term, which is always real, symmetric, and positive-definite. However, in our case, $\text{sym}\Big(FS_{ij}^{LR}\Big)$ is, in general, complex-valued, much like indefinite metric spaces similar to Lorentzian geometry.

At this point, it is worth mentioning that, in the Hermitian paradigm, the dynamics governed by the quantum natural gradient descent is equivalent to the imaginary-time evolution problem in the vanishing step size limit \cite{Stokes}. On the other hand, since in the biorthogonal formalism, the dynamics of the left and right states are governed by two generators in the generic case, which are not Hermitian adjoint of each others, there is no single tangent direction of the flow (indicated by the presence of the two step sizes above in the generic situations), the natural gradient algorithm with the cost determined by the Hamiltonian (for the Hermitian case) is not equivalent to the imaginary-time evolution, even after imposing the norm conservation condition during the evolution. However, if we want to optimise a cost constructed from a right state, and the corresponding `associated' left state, then it forces a single flow direction, and the imaginary-time evolution problem indeed reduces to the quantum natural gradient problem, governed by the symmetric part of the LR (RL) FS tensor.
We discuss this issue briefly in the appendix \ref{Appendix2}, where we show why the straightforward generalisation of the imaginary-time evolution problem with the aim of minimising the overlap between an evolved state and that state projected on the variational submanifold fails to be a proper minimisation problem in the general biorthogonal formalism.

\subsubsection{LR (RL) connections and their duality}
In this section, we will briefly discuss the connections induced on the parameter manifold from the non-Hermitian inner product structure for $\alpha=0$. Following similar expansion as we did in the section \ref{nmconnections1}, we can obtain the two rank-3 non tensorial connections from the expansion of the overlap integral $\mathcal{D}^{LR}(\theta, \theta^{\prime})$, which are not gauge invariant, and of the form, 
\begin{equation}
\tilde{\Gamma}_{ij,k}^{LR}=\braket{\partial_{i}\partial_{j}{\Psi^{L}}|\partial_{k}\Psi^R}, ~~\text{and}~~ \tilde{\Gamma}_{ij,k}^{RL}=\braket{\partial_{k}{\Psi^{L}}|\partial_{i}\partial_{j}\Psi^R}~.
\label{LRconnections}
\end{equation}
These two connections are dual with respect to each other under the simultaneous $L \rightarrow R$ and complex conjugation operation, which is a reduced version of the $**$ duality we mentioned in \eqref{generaliseduality} as  we are dealing with only $\alpha=0$ case. Note that in the classical information geometry, the $\alpha=0$ case is the metric connection of the FR metric. The physically meaningful, gauge-invariant connections can be obtained from  these two in \eqref{LRconnections} are of the form 
\begin{widetext}
\begin{equation}
\begin{split}
\Gamma_{ij,k}^{LR}=\braket{\partial_{i}\partial_{j}{\Psi^{L}}|\partial_{k}\Psi^{R}} -\Big(\braket{\partial_{i}\partial_{j}{\Psi^{L}}|\Psi^{R}}\braket{{\Psi^{L}}|\partial_{k}\Psi^{R}}+2\braket{\partial_{(i}{\Psi^{L}}|\Psi^{R}}\braket{\partial_{j)}{\Psi^{L}}|\partial_{k}\Psi^{R}}\Big)
+2\braket{\partial_{i}{\Psi^{L}}|\Psi^{R}}\braket{\partial_{j}{\Psi^{L}}|\Psi^{R}}\braket{{\Psi^{L}}|\partial_{k}\Psi^{R}}~, 
\label{alphaconnectiongauged}   
\end{split}
\end{equation}
\end{widetext}
and 
\begin{widetext}
    \begin{equation}
\begin{split}
\Gamma_{ij,k}^{RL}=\braket{\partial_{k}{\Psi^{L}}|\partial_{i}\partial_{j}\Psi^{R}}-\Big(\braket{{\partial_{k}\Psi^{L}}|\Psi^{R}}\braket{{\Psi^{L}}|\partial_{i}\partial_{j}\Psi^{R}}+2\braket{{\Psi^{L}}|\partial_{(i}\Psi^{R}}\braket{{\partial_{k}\Psi^{L}}|\partial_{j)}\Psi^{R}}\Big)
+2\braket{\partial_{k}{\Psi^{L}}|\Psi^{R}}\braket{{\Psi^{L}}|\partial_{i}\Psi^{R}}\braket{{\Psi^{L}}|\partial_{j}\Psi^{R}}~.
\label{alphamconnection}
\end{split}
\end{equation}
\end{widetext}
Note that due to the non-Hermitian, biorthogonal inner product structure, even for the $\alpha=0$ case, these two connections $\Gamma_{ij,k}^{LR}$ and $\Gamma_{ij,k}^{RL}$ will satisfy a non-trivial transformation duality, that we call left-right duality of these two connections, which is trivial in the classical or in the case of connection induced from the standard Hermitian inner product. This can also be seen from the fact that, instead of expanding  $\mathcal{D}^{LR}(\theta, \theta^{\prime})$, we can, in principle, also work with $\mathcal{D}^{RL}(\theta, \theta^{\prime})=\braket{\Psi^{R}(\theta+\delta\theta)-\Psi^{R}(\theta)|\Psi^{L}(\theta+\delta\theta)-\Psi^{L}(\theta)}$, with proper choice of normalisation, then we will obtain complex conjugates of $\Gamma_{ij,k}^{RL}$ and $\Gamma_{ij,k}^{LR}$ from the expansion in the \textit{reverse} order.

\subsection{LL(RR)-Fubini-Study tensor}
\label{leftleftFStensorsection}
Let us now comment on the case where the `metric' is constructed from either only the left or only the right states. First of all, this kind of construction is valid only when the normalisation of the states is such that 
\begin{equation}
    \braket{\Psi^{I}|\Psi^{I}}=1,
\label{leftorrightnorm}
\end{equation}
for $I=L, R$, provided they can be normalised, which is not guaranteed for a general non-hermitian operator, and it is not meaningful to compare this inner product with the biorthogonal inner product with respect to which we have defined \eqref{leftrightnorm}. It is also possible to construct the geometry of LL (or RR) states without enforcing the pointwise normalisation as we implemented above, but the geometry in that case will typically not be on the projective space.  Note also that even though we have continued to use $\bra{\Psi^{I}}$ to denote the Hermitian conjugate of $\ket{\Psi^{I}}$, this should not be confused with $\bra{\Psi^{I}}_{\mathcal{B}}$, which can be thought of as the dual with respect to the biorthogonal pairing, and was called the associated state of $\ket{\Psi}$ in \cite{Brody2013biorthogonal}. Consequently, the LR FS tensor in section \ref{LRFStensorsection} was  constructed by considering two nearby states $\ket{\Psi^{R}(\theta+\delta\theta)}$ and $\ket{\Psi^{R}(\theta)}$, where the associated state of $\ket{\Psi^{R}(\theta+\delta\theta)}$ was `constructed' respecting the biorthogonal conjugation, which is $\bra{\Psi^{L}(\theta)}_{\mathcal{B}}$ and $\bra{\Psi^{L}(\theta+\delta\theta)}_{\mathcal{B}}$ respectively.  Hence, the geometric tensors in the corresponding cases are `distances' with respect to completely differ inner products, and  it is not consistent to compare the tensor constructed from either left (or right only) states $FS^{LL}$ (or $FS^{RR}$) and that from using both left \textit{and} right tensors $FS^{LR}$ (or $FS^{RL}$) in the same setting together. 


To reiterate, the correct strategy will be to fix the inner product and the normalisation of the corresponding states first and then construct the overlap of nearby states to extract the metric tensor, as the normalisation is a crucial ingredient in this process. On the other hand, as we will explain, $FS^{II}$ and $FS^{LR}$ essentially capture two somewhat different aspects of state space geometry, and the physical information they encode will be, in general, different. Here we have to consider the overlap of two nearby left (or right) states, where the conjugate-linear inner product is defined with respect to only left (or right) states as
$\braket{\Psi^{I}(\theta+\delta\theta)-\Psi^{I}(\theta)|\Psi^{I}(\theta+\delta\theta)-\Psi^{I}(\theta)}$. Then using the normalisation \eqref{leftorrightnorm}, we arrive at the following. 
\begin{equation}
FS_{ij}^{II}=\braket{\partial_{i}\Psi^{I}|\partial_{j}\Psi^{I}}-\braket{\partial_{i}\Psi^{I}|\Psi^{I}}\braket{\Psi^{I}|\partial_{j}\Psi^{I}}~. 
\label{leftorrighttensor}
\end{equation}
To obtain this, we impose the gauge-invariance of this $2$-form, which is of the form $\ket{\Psi^{I}} \rightarrow e^{i\beta_{I}}\ket{\Psi^{I}}$, which also fixes the transformation of the conjugate vector $\bra{\Psi^{I}}$. Note that the tensor structure \eqref{leftorrightnorm} is essentially Hermitian (although it has important differences from the standard Hermitian QGT; footnote 9), and it has \textit{only} two components, the standard real and symmetric part (QMT) and the Berry curvature; all the flipped tensors introduced above vanish identically,
\begin{equation}
    g_{ij}^{II}=\frac{1}{2}\Big[(FS_{ij}^{II})+(FS_{ji}^{II})\Big]~,\hspace{2mm}  \text{and} \hspace{2mm}   \omega_{ij}^{II}=\frac{1}{2}\Big[(FS_{ij}^{II})-(FS_{ji}^{II})\Big]~.
\label{metricleftorright}
\end{equation} 

Intrinsically, these tensors live on the complex projective space \footnote{It is important to note that, even though it is possible to consider a complex projective space when we are working with a point wise parameter-dependent normalised state, the projection via $U(1)$ phase freedom, it is \textit{not} generated by any group action, and consequently \textit{not} invariant under dynamical evolution, which will not be in the projective equivalence class. Non-unitary transformations are not an isometry of such a space, a fact  reflected in the non-Hermitian tensor structure, that is not invariant under evolution, which we discuss briefly  in appendix \ref{Appendix3}.} of either the left or the right vector spaces and are then pulled back to the parameter manifold, which is equipped with the coordinate charts defined by the parameters of the system. The inner product in this case is a map from the left (or right) vector space $\mathcal{V}_{L}$ and its dual $\mathcal{V}_{L}^{*}$ to the space of complex numbers (provided that the conjugation map exists), which is fundamentally different from the bilinear inner product for left \textit{and} right states in $(FS)^{LR}$. Consequently, the two real and symmetric rank-2 tensors $g_{ij}^{II}$ and $g_{ij}^{LR}$ capture `distance' in two different settings, and it is not meaningful to compare these two. Furthermore, the LL(RR) FS tensors, though are `Hermitian', but are not identical in effect to that of the geometry of standard Hermitian quantum systems, as on the parameter manifold, the left states $\ket{\Psi_{L}}$ (or the right $\ket{\Psi_{L}}$ ) and the corresponding complex conjugates $\bra{\Psi_{L}}$ do not belong to the unitary orbit of $\Psi_{L}(\theta=0)$ (or $\Psi_{R}(\theta=0)$)  like that of the Hermitian counterpart. Consequently, the LL (or RR) QMT may not be positive definite always, particularly near the exceptional points.

\subsubsection{Optimisation problem and the RR (LL) QMT}
\label{optimisation2}
The difference between the complex-valued LR (RL)-QMT and the LL(RR)-QMT can probably be best illustrated by considering the natural gradient optimisation problem, where in this case the cost is tailored for right-right or left-left states. As was already described above in section \ref{optimisation2}, for the non-hermitian systems, setting up a variational eigenvector problem in the traditional way is not possible, as the eigenvalue of the Hamiltonian might be complex-valued. However, recent works have used the energy variance of such Hamiltonian as a cost to perform a similar variational quantum algorithm \cite{Guo22}. To this end, one typically assigns a Hermitian counterpart of a non-Hermitian Hamiltonian in such a way that the cost associated with this Hermitian counterpart can be used to run a conventional variational algorithm with proper modification. Then, for a non-Hermitian Hamiltonian $\mathcal{H}$, we have the Hermitian counterpart constructed from the right eigenstates of $\mathcal{H}$,
\begin{equation}
    \tilde{\mathcal{H}}^{RR}=\Big(\mathcal{H}^{\dagger}-(\bar{E}^{R})\Big)\Big(\mathcal{H}-(E^{R})\Big)~,
\end{equation}
where $E^{R}$, is the right eigenvalue, and $\bar{E}^{R}$ is the complex conjugate. The optimisation condition for the Hermitian counterpart, 
\begin{equation}
    \mathcal{L}^{RR}=\braket{\Psi^{R}(\theta)|\tilde{\mathcal{H}}^{RR}|\Psi^{R}(\theta)}
\label{RRcost}
\end{equation}
for a parametrised set of right states $\Psi^{R}(\theta)$ will then imply that the optimised state is actually a right eigenstate. This kind of proposal was used, for example, in \cite{Xie24} to determine the right and left eigenstates of non-Hermitian Hamiltonians. In our context, the real and symmetric RR-QMT is the natural indicator of the optimisation direction for this type of cost \eqref{RRcost}, in the natural gradient descent approach, for the effective Hermitian operator $\tilde{\mathcal{H}}^{RR}$, which can be thought to be a non-Hermitian generalisation of the proposal of \cite{Stokes}. Then a local solution of the optimisation problem is, 
\begin{equation}
   g_{ij}^{RR}\delta\theta^{j}=-\eta^{RR}\partial_{i}\mathcal{L}^{RR}~,  \end{equation}
such that the optimal direction on the parameter manifold is 
\begin{equation}
    \theta_{t+1}^{i}=\theta^{i}_{t}-\eta^{RR} g^{(RR)ij}\partial_{j}\mathcal{L}^{RR}(\theta)~.
\end{equation}
This determines a unique solution of the optimisation problem (as long as $g_{ij}^{RR}$ is invertible, otherwise the pseudo-inverse has to be used), and moves the point $\theta_{i}$ to the opposite direction of the gradient of $\mathcal{L}^{RR}(\theta)$ for the underlying metric of the right-right state space, $g_{ij}^{RR}$.

\subsubsection{LL (RR) connections}
For the sake of completeness, we also briefly mention the connections built solely from LL (or RR) overlap integral, $\mathcal{D}^{II}(\theta, \theta^{\prime})=\braket{\Psi^{I}(\theta+\delta\theta)-\Psi^{I}(\theta)|\Psi^{I}(\theta+\delta\theta)-\Psi^{I}(\theta)}$, which will of the form, after restoring gauge invariance,
\begin{widetext}
\begin{equation}
\begin{split}
\Gamma_{ij,k}^{II}=\braket{\partial_{i}\partial_{j}{\Psi^{I}}|\partial_{k}\Psi^{I}} -\Big(\braket{\partial_{i}\partial_{j}{\Psi^{I}}|\Psi^{I}}\braket{{\Psi^{I}}|\partial_{k}\Psi^{I}}+2\braket{\partial_{(i}{\Psi^{I}}|\Psi^{I}}\braket{\partial_{j)}{\Psi^{I}}|\partial_{k}\Psi^{I}}\Big)
+2\braket{\partial_{i}{\Psi^{I}}|\Psi^{I}}\braket{\partial_{j}{\Psi^{I}}|\Psi^{I}}\braket{{\Psi^{I}}|\partial_{k}\Psi^{I}}~, 
\label{alphaconnectiongaugedII}   
\end{split}
\end{equation}
\end{widetext}
and 
\begin{widetext}
    \begin{equation}
\begin{split}
\bar{\Gamma}_{ij,k}^{II}=\braket{\partial_{k}{\Psi^{I}}|\partial_{i}\partial_{j}\Psi^{I}}-\Big(\braket{{\partial_{k}\Psi^{I}}|\Psi^{I}}\braket{{\Psi^{I}}|\partial_{i}\partial_{j}\Psi^{I}}+2\braket{{\Psi^{I}}|\partial_{(i}\Psi^{I}}\braket{{\partial_{k}\Psi^{I}}|\partial_{j)}\Psi^{I}}\Big)
+2\braket{\partial_{k}{\Psi^{I}}|\Psi^{I}}\braket{{\Psi^{I}}|\partial_{i}\Psi^{I}}\braket{{\Psi^{I}}|\partial_{j}\Psi^{I}}~.
\label{alphamconnectionII}
\end{split}
\end{equation}
\end{widetext}
Here, since $FS^{II}$ is essentially Hermitian, the duality of the LR connections is even further reduced, and the only remaining one dictates that two connections are complex conjugates of each other, and can be thought of as non-Hermitian extensions of the connections in \cite{Heteneyi}.

\section{Conclusions and Discussion}
The duality of the Fisher-Rao metric and the two $\pm\alpha$ connections is of central importance in classical information geometry. When dealing with quantum systems, there are various rigorous generalisations of such results in the quantum information geometry formalism, primarily assuming a Hermitian inner product structure on the Hilbert space. In this work, our primary motivation is to investigate how the modification of the standard tensor structure, obeying the Hermitian  product affects these results on the complex projective manifold of pure quantum states, which is naturally equipped with the Hermitian Fubini-Study metric induced by the Hermitian inner product of the underlying Hilbert space, which can be written as a Fisher-Rao metric for the probability distribution, modified by the presence of the non-trivial quantum phases. 

As emphasised in the main text, one of the most natural generalisations of such duality is to consider a non-Hermitian inner product, and consequently a non-Hermitian, biorthogonal extension of the standard Fubini-Study metric. This gives rise to a rich structure when pulled back to the parameter submanifold of interest, which includes, in particular, four components of the Fubini-Study tensor, consisting of two flipped (real antisymmetric and imaginary symmetric tensors) apart from the standard form of the quantum metric tensor and the Berry curvature. These flipped tensors, in general, affect the metric and the Berry curvature, both of which are generally complex-valued in the non-Hermitian case. We provided a systematic formulation of the Fubini-Study tensor and the two associated connections from an expansion of the overlap integral and showed how the real part of these two gauge-invariant connections indeed satisfies a $\alpha$-like duality in a generalised way, taking the effects of quantum fluctuations. To understand the role of each component of this non-Hermitian Fubini-Study tensor, we considered the quantum natural gradient flow and showed that, for a general complex-valued loss function, the real, symmetric, and imaginary, symmetric parts of the non-Hermitian Fubini-Study tensor provide mutually non-compatible directions of optimisation for the real parameter manifold.

When the quantum system is governed by a non-Hermitian Hamiltonian, it is possible to consider different state space geometries, depending on the context, and they are, in general, not equivalent and are expected to capture different aspects of the geometry of the parameter manifold induced by different inner products on the Hilbert space. Starting with the overlap integral in a biorthogonal formalism, we have provided a systematic classification of the four kinds of tensors that might arise in a left-right, right-left, left-left and right-right formalism. We have elucidated the important role of the normalisation chosen in each case, and classified the most generic non-Hermitian tensor into four categories: real and symmetric, purely imaginary and symmetric, real and antisymmetric, and purely imaginary and antisymmetric components.  The metric that governs the natural gradient in the parameter submanifold then receives contributions from the purely imaginary and symmetric tensor, apart from the standard real and symmetric quantum metric tensor. On the other hand, the complex-valued Berry curvature, which is the curvature tensor of the non-Hermitian Berry connection, is then composed of the real and antisymmetric parts also along with the standard purely imaginary contributions. We have also emphasised the difference between constructing tensors composed of left-right (right-left) states and those of left-left (right-right) states, as they are fundamentally different, with different ranges of validity as well and should be used depending on the geometry of the observable we are interested in.  

We  conclude with a brief discussion on the several important features of the formalism used in this work that need to be addressed further and the various directions in which we believe our results can be used. First of all, we have pointed out a generalised duality that exists between the three complex-valued objects, the Fubini-Study tensor, and the two $\pm$ connections, that resembles the corresponding duality in the classical information geometry. We can, of course, interpret the real part of our result in a similar way to that of the existence of affine coordinates, as long as we are using only the real parts of these objects. However, two points have to be noted before such an interpretation; the first one is that, unlike the classical probability distribution, for quantum systems, apart from the probability amplitude, the phase (even though it is not observable in the standard single-particle quantum mechanics)  is expected to play a crucial role, and secondly, we have seen how the purely imaginary, but symmetric part of the Fubini-Study tensor also is important for the flow of a general complex-valued cost, it will require further careful investigation about proper duality on the full complex projective space. We have also seen that if the natural gradient descent problem in the generic non-Hermitian cost function gives two mutually incompatible solutions of the flow directions, then the natural question to ask is whether it is possible to obtain a mutually compatible solution for this overdetermined system. It might be possible to solve this kind of system using least-squares techniques for overdetermined systems, and we hope to report this in the near future.

\section{Acknowledgments}
It is a pleasure to thank Mir Afrasiar, Arpita Mitra, Debangshu Mukherjee and Kuntal Pal for several discussions and for comments on a draft version of the manuscript. The work of KP is  supported by the Young Scientist Training Program  at the APCTP through the Science and
Technology Promotion Fund and the Lottery Fund of the Korean Government. This was also supported by the Korean Local Governments -
Gyeongsangbuk-do Province and Pohang City.

\appendix
\section{Condition for the validity of the biorthogonal structure throughout the `evolution'}
\label{Appendix1}
In this section, we will briefly discuss the condition that the generators of the $s$-evolutions discussed in sec. \ref{inducedfluctuation} have to be satisfied for the biorthogonal structure to be preserved during evolution $s$. By this, we mean if $\braket{l_{1(\alpha)}|l_{2(-\alpha)}}=1$ for all $s$, if we start with $\braket{l^{0}_{1(\alpha)}|l^{0}_{2(-\alpha)}}=1$ at $s=0$. To this end, we define $f(s)=\braket{l_{1(\alpha)}(s)|l_{2(-\alpha)}(s)}$, as a continuous function of $s$, which interpolates from $s=0$ to an arbitrary value of $s$. For consistency of the biorthogonality condition, we require $f^{\prime}=0$ at any arbitrary point on the evolution curve, where the prime denotes the derivative of the function with $s$. Taking the derivative with respect to the parameter $s$, it is evident that the condition reduces to the constraints of $Z(s)\in Ker(u)$, where we have defined $Z(s)=M(s)\Big(A_{2(-\alpha)}-M(s)^{-1}(A_{1(\alpha)})^{\dagger}M(s)\Big)v$. Here, for simplicity, we have assumed that the states depend on a single parameter $s$, and $u$, and $v$ are the (matrix representations) of the left and the right initial states, respectively.

\section{Imaginary-time evolution in the biorthogonal formalism}
\label{Appendix2}
In this section, we will briefly describe why the  natural gradient descent optimisation and the imaginary-time evolution for the evolution generated by the non-Hermitian generators, where for the Hermitian counterparts, they are, see, for example, \cite{Stokes}. In particular, we show that, even though they are equivalent when, using the biorthogonal structure and the associated left and right states, it might not be so when dealing with only right (or left) states and the corresponding Hermitian conjugates. 
We will primarily emphasise on two points, why the Hermitian nature of the evolution generators, both under the biorthogonal and the standard inner product is necessary for the above equivalence to be valid, for a well-defined cost function, when the time-evolved state is projected back to the parameter submanifold, and how the variational condition choses the symmetric part of the non-Hermitian FS tensor (which might not be real), providing a clear interpretation of such quantities as complex valued metric.
\subsection{LR (or RL) imaginary-time evolution}
Let us consider the following imaginary-time evolution of the biorthogonal right state and the corresponding associated state combinations generated respectively by $H_{L}$ and $H_{R}$, such that the time-evolved states are 
\begin{equation}
\begin{split}
  \ket{\Psi_{R}(\tau;\theta)}=e^{-H_{R}\delta\tau}\ket{\Psi_{R}(\tau=0;\theta)}~, 
   ~\text{and}~ \\
   \bra{\Psi_{L}(\tau;\theta)}=\bra{\Psi_{L}(\tau=0;\theta)}e^{-H^{\#}_{R}\delta\tau}=\bra{\Psi_{L}(\tau=0;\theta)}e^{-H_{R}\delta\tau}, 
\end{split}
   \end{equation}
where we have assumed that the states at $\tau=0$ are $\bra{\psi_{L}(\tau=0;\theta)}$ and $\ket{\psi_{R}(\tau=0;\theta)}$, and $H^{\#}_{R}$ is the `Hermitian adjoint' of the operator $H_{R}$. Since we are considering the evolution of a right state $\ket{\Psi_{R}}$ and the related associated state $\bra{\Psi_{L}}$ \cite{Brody2013biorthogonal}, the generator is biorthogonal-Hermitian $H^{\#}_{R}=H_{R}$. The time-evolved states are then projected back to the parameter submanifold at $\theta+\delta\theta$ so that the difference between the (imaginary) time-evolved state and the projected state is minimised 
\begin{equation}
    \text{min}\Bigg(\Big(\bra{\Psi_{L}(\tau;\theta)}-\bra{\Psi_{L}(\tau;\theta)}\mathcal{P}^{LR}_{\theta+\delta\theta}\Big)\Big(\ket{\Psi_{R}(\tau;\theta)}-\mathcal{P}^{LR}_{\theta+\delta\theta}\ket{\Psi_{R}(\tau;\theta)}\Big)\Bigg),
\end{equation}
for real-valued parameter increment $\delta\theta$. Expanding this overlap up to quadratic order in increments (both for time and parameters), we have the following
\begin{widetext}
 \begin{equation}
\delta\theta^{i}\delta\tau\Big(\braket{\Psi_{L}|H_{R}|\partial_{i}\Psi_{R}}+\braket{\partial_{i}\Psi_{L}|H_{R}|\Psi_{R}}\Big)+\delta\theta^{i}\delta\theta^{j}\Big(\braket{\partial_{i}\Psi^{L}|\partial_{j}\Psi^{R}}-\braket{\partial_{i}\Psi^{L}|\Psi^{R}}\braket{\Psi^{L}|\partial_{j}\Psi^{R}}\Big)~,
\end{equation}   
\end{widetext}
where we have used the biorthogonal normalisation condition.
Then the optimisation condition gives us
 \begin{equation}
\delta\tau\Big(\braket{\Psi_{L}|H_{R}|\partial_{i}\Psi_{R}}+\braket{\partial_{i}\Psi_{L}|H_{R}|\Psi_{R}}\Big)+\delta\theta^{j}\Big(\text{Sym}(FS^{LR}_{ij})\Big)=0~,
\end{equation} 
where only the symmetric part of the LR-FS tensor, $\text{Sym}(FS^{LR}_{ij})$ contributes to the optimisation problem. This shows that, even though the kinematics of the evolution projected in the parameter space is always governed by the symmetric but complex-valued part of the left-right FS tensor, the imaginary-time evolution is not equivalent to the gradient descent optimisation if $H^{\#}_{L}\neq H_{R}$. i.e., not biorthogonal-Hermitian. On the other hand, in the biorthogonal-Hermitian case $H^{\#}_{L}= H_{R}$, the optimisation problem reduces to 
\begin{equation}
   \frac{\partial\theta^{i}}{\partial\tau}=-\text{Sym}(FS^{(LR)ij})\frac{\partial \mathcal{L}}{\partial\theta^{j}}~,
\end{equation}
for the cost functional $\mathcal{L}(\theta)=\braket{\Psi_{L}|H_{R}|\Psi_{R}}$. In the standard Hermitian limit, this exactly reduces to the natural gradient optimisation for the symmetric (and real) part of the Hermitian FS tensor \cite{Stokes}.

\subsection{RR (or LL) imaginary-time evolution}
Let us now consider the imaginary-time evolution of only the  right-right  (or the left-left) state combinations $\ket{\Psi_{R}}, \bra{\Psi_{R}}$ (or $\ket{\Psi_{L}}, \bra{\Psi_{L}}$) generated respectively by $H_{R}$ and $H^{\dagger}_{R}$ (or $H_{L}$ and $H^{\dagger}_{L}$), such that the time-evolved states are 
\begin{equation}
\begin{split}
   \ket{\Psi_{R}(\tau;\theta)}=e^{-H_{R}\delta\tau}\ket{\Psi_{R}(\tau=0;\theta)}~, ~\text{and}~ \\ 
   \bra{\Psi_{L}(\tau;\theta)}=\bra{\Psi_{L}(\tau=0;\theta)}e^{-H^{\dagger}_{R}\delta\tau},
\end{split}
   \end{equation}
where we have assumed that the states at $\tau=0$ are $\bra{\psi_{R}(\tau=0;\theta)}$ and $\ket{\psi_{R}(\tau=0;\theta)}$. The time-evolved states are then projected back to the parameter submanifold at $\theta+\delta\theta$ so that the difference between the (imaginary) time-evolved state and the projected state is minimised, 
\begin{equation}
    \text{min}\Bigg(\Big(\bra{\Psi_{R}(\tau;\theta)}-\bra{\Psi_{R}(\tau;\theta)}\mathcal{P}^{RR}_{\theta+\delta\theta}\Big)\Big(\ket{\Psi_{R}(\tau;\theta)}-\mathcal{P}^{RR}_{\theta+\delta\theta}\ket{\Psi_{R}(\tau;\theta)}\Big)\Bigg),
\end{equation}
for real-valued parameter increment $\delta\theta$. Expanding this overlap up to quadratic order in increments (both for time and parameters), we have the following
\begin{widetext}
 \begin{equation}
\delta\theta^{i}\delta\tau\Big(\braket{\Psi_{R}|H^{\dagger}_{R}|\partial_{i}\Psi_{R}}+\braket{\partial_{i}\Psi_{R}|H_{R}|\Psi_{R}}\Big)+\delta\theta^{i}\delta\theta^{j}\Big(\braket{\partial_{i}\Psi^{R}|\partial_{j}\Psi^{R}}-\braket{\partial_{i}\Psi^{R}|\Psi^{R}}\braket{\Psi^{R}|\partial_{j}\Psi^{R}}\Big)~,
\end{equation}   
\end{widetext}
where we have used two conditions: the right-right normalisation condition and the requirement that this norm will be conserved throughout the real-time evolution. Note also that, in particular for RR (or LL) state evolution on the parameter space, we have to use only the point-wise normalised states (see \ref{Appendix3}, states like $\tilde{\Psi}_{R}(\theta)$) only otherwise all the projection operations will have to normalise separately.
Then the optimisation condition gives us
 \begin{equation}
\delta\tau\Big(\braket{\Psi_{R}|H^{\dagger}_{R}|\partial_{i}\Psi_{R}}+\braket{\partial_{i}\Psi_{R}|H_{R}|\Psi_{R}}\Big)+\delta\theta^{j}\Big(\text{Sym}(FS^{RR}_{ij})\Big)=0~,
\end{equation} 
where  the real and symmetric part of the RR-FS tensor, $g^{RR}_{ij}$ contributes to the optimisation problem. This shows that, even though the kinematics of the evolution projected in the parameter space is governed by the symmetric but complex-valued part of the left-right FS tensor, the imaginary-time evolution is not equivalent to the gradient descent optimisation if $H^{\dagger}_{R}\neq H_{R}$. To summarise, the LR imaginary-time evolution corresponds to a variational scheme that tracks a decay direction in the complex energy landscape preserving the biorthogonal structure, and is natural in a variational setting. RR evolution, in contrast, does not preserve a variational cost in the same way unless the generator is Hermitian. Hence, the optimisation direction is not guaranteed to correspond to steepest descent. Whether that can be done by modifying the normalisation or the definition of the cost functional remains to be seen.


\section{Geometry of Norm-invariant RR (or LL) states from QFI}
\label{Appendix3}
Since the norm of a reference state is not invariant under the action of non-Hermitian generator, unless specific conditions are imposed, it is important to properly normalise the `evolved' state at each instant of the trajectory to obtain a well-defined tensor structure on the projective space. Consequently, the explicit form of the Fubini-Study tensor will be different for a properly norm-preserving state and another state, where the norm changes during the evolution. Let us consider a parametrised family of states $\Psi_{I}(\theta)$, generated by a non-Hermitian operator $\mathcal{H}_{I}$, from a reference state $\Psi_{I}(\theta=0)$, 
\begin{equation}
    \ket{\Psi_{I}(\theta)}=U_{\theta}\ket{\Psi_{I}(\theta=0)}~,
\end{equation}
with $U_{\theta}=e^{-i\mathcal{H}_{I}\theta}$. Then for a generic non-Hermitian operator $\mathcal{H}_{I} \neq \mathcal{H}_{I}^{\dagger}$, the state $\Psi_{I}(\theta)$ will not be normalised, for $\braket{\Psi_{I}(\theta=0)|\Psi_{I}(\theta=0)}=1$. However, it is still meaningful to define a point-wise (dependent on $\theta$) normalised state as
\begin{equation}
\ket{\tilde{\Psi_{I}}(\theta)}=\frac{\ket{\Psi_{I}(\theta)}}{\braket{\Psi_{I}(\theta)|\Psi_{I}(\theta)}}=\frac{1}{N(\theta)}U_{\theta}\ket{\Psi_{I}(\theta=0)}~,
\end{equation}
with the parameter-dependent normalisation factor
$N^{2}(\theta)=\braket{\Psi_{I}(\theta=0)|U_{\theta}^{\dagger}U_{\theta}|\Psi_{I}(\theta=0)}$. Then the FS tensors for this set of normalised states are of the form 
\begin{equation}
FS^{II}_{ij}=\braket{\tilde{\Psi_{I}}(\theta)|\tilde{\Psi_{I}}(\theta)}-\braket{\partial_{i}\tilde{\Psi_{I}}(\theta)|\tilde{\Psi_{I}}(\theta)}\braket{\tilde{\Psi_{I}}(\theta)|\partial_{j}\tilde{\Psi_{I}}(\theta)}~,
\end{equation}
which, when written in terms of the states without normalisation assumes the form 
\begin{equation}
FS^{II}_{ij}=\frac{1}{N^2(\theta)}\Big(\braket{\Psi_{I}(\theta)|\Psi_{I}(\theta)}-\frac{1}{N^2(\theta)}\braket{\partial_{i}\Psi_{I}(\theta)|\Psi_{I}(\theta)}\braket{\Psi_{I}(\theta)|\partial_{j}\Psi_{I}(\theta)}\Big)~,
\end{equation}
 which essentially captures the explicit normalisation factors, which are necessary to capture the effect parameter dependent norms and must be included when considering the geometry of these kinds of states.   Then with respect to the normalised density matrix $\tilde{\rho}^{I}(\theta)=\ket{\tilde{\Psi}^{I}(\theta)}\bra{\tilde{\Psi}^{I}(\theta)}=\frac{1}{N^2(\theta)}U_{\theta}\ket{\Psi_{I}(\theta=0)}\bra{\Psi_{I}(\theta=0)}U^{\dagger}_{\theta}$, which is essentially Hermitian, we can construct the symmetric logarithmic derivative, and define the QFI as 
 \begin{equation}
\mathcal{F}_{I}=4\text{Tr}\Big[\tilde{\rho}^{I}(\theta)\partial_{i}\tilde{\rho}^{I}(\theta)\partial_{j}\tilde{\rho}^{I}(\theta)\Big]~.
 \end{equation}
Then, evaluating the trace and utilising the fact for the normalised density matrix, the trace of the density is preserved throughout the evolution, we obtain, 
\begin{widetext}
  \begin{equation}
 \mathcal{F}_{I}= \frac{4}{N^2(\theta)}\Big(\braket{\Psi_{I}(\theta=0)|\partial_{i}U^{\dagger}_{\theta}\partial_{j}U_{\theta}|\Psi_{I}(\theta=0)}-\frac{1}{N^2(\theta)}\braket{\Psi_{I}(\theta=0)|\partial_{i}U^{\dagger}_{\theta}U_{\theta}|\Psi_{I}(\theta=0)}\braket{\Psi_{I}(\theta=0)|U^{\dagger}_{\theta}\partial_{j}U_{\theta}|\Psi_{I}(\theta=0)}\Big)~,  
\end{equation}
\end{widetext}
where we have used $\partial_{i}\Big(\frac{1}{N^2(\theta)}\braket{\Psi_{I}(\theta=0)|U^{\dagger}_{\theta}U_{\theta}|\Psi_{I}(\theta=0)}\Big)=0$. This reduces to \cite{Yu2023}
\begin{equation}
\mathcal{F}_{I}=4\Big(\braket{\tilde{\Psi_{I}}(\theta)|\mathcal{H}_{I}^{\dagger}\mathcal{H}_{I}|\tilde{\Psi_{I}}(\theta)}-\braket{\tilde{\Psi_{I}}(\theta)|\mathcal{H}_{I}^{\dagger}|\tilde{\Psi_{I}}(\theta)}\braket{\tilde{\Psi_{I}}(\theta)|\mathcal{H}_{I}|\tilde{\Psi_{I}}(\theta)}\Big)~.
\end{equation}
When the evolution is unitary, generated by a Hermitian operator, QFI is the variance of the generator, $4\braket{\Psi_{0}|(\Delta \mathcal{H})^2|\Psi_{0}}$, and depends only on the initial state, not on the particular trajectory on the parameter manifold. On the other hand, for the evolution generated by the non-Hermitian operator, the QFI of a parametrised state explicitly depends on the path traversed by the mapping $\tilde{\Psi_{I}}(\theta)$.

\bibliography{reference}
\end{document}